\documentclass{aa}  

\usepackage{graphicx}
\usepackage{array}
\usepackage{amsmath}
\usepackage{txfonts}
\usepackage{caption}
\usepackage{lipsum}
\usepackage[dvipsnames]{xcolor}
\usepackage{hyperref}
\usepackage{url}
\usepackage{natbib}
\usepackage{breqn}
\usepackage{diagbox}
\usepackage{cuted} 
\usepackage[mathscr]{euscript}
\usepackage{adjustbox}
\usepackage{multirow}
\usepackage[nice]{nicefrac}
\usepackage{tabularx}
\usepackage{orcidlink}

\DeclareMathOperator\erf{erf}
\newcolumntype{C}{>{\centering\arraybackslash}p{0.2\textwidth}}
\usepackage{tablefootnote} 
\usepackage{enumitem}
\usepackage{booktabs}
\bibpunct{(}{)}{;}{a}{}{,} % to follow the AA style
\begin{document}

%From Embedded Ionization to Expanding H{\sc\,ii} Regions: Predicting Dust and Free–Free Continuum Emission

\title{Predicting Continuum Emission in Massive Star-Forming Regions with Embedded H{\sc\,ii} Regions - Dust, Free-Free, and Anomalous Microwave Emission with POLARIS}

 \titlerunning{Predicting Continuum Emission in Massive Star-Forming Regions}
 
\author{Stefan Reissl \inst{\ref{inst1}}\orcid{0000-0001-5222-9139} \and   Alessio Traficante
  \inst{\ref{inst2}}\orcidlink{0000-0003-1665-6402} \and Paolo Suin \inst{\ref{inst3}, \ref{inst8}}\orcid{0000-0001-7044-3809} \and Alice Nucara \inst{\ref{inst2},\ref{inst4}}\orcidlink{0009-0005-9192-5491} \and Ralf S. Klessen \inst{\ref{inst1},\ref{inst5}}\orcid{0000-0002-0560-3172} \and Leonardo Testi \inst{\ref{inst6},\ref{inst7}}\orcid{0000-0003-1859-3070} % \and others yet to come \inst{\ref{inst1}}
}
\institute{
\centering Universit\"{a}t Heidelberg, Zentrum f\"{u}r Astronomie, Institut f\"{u}r Theoretische Astrophysik, Albert-Ueberle-Straße 2, \\  D-69120 Heidelberg, Germany \label{inst1} 
\and
\centering Istituto Nazionale di Astrofisica (INAF)-Istituto di Astrofisica e Planetologia Spaziali, Via Fosso del Cavaliere 100,
I-00133 Roma, Italy \label{inst2}
\and
\centering Aix Marseille Univ, CNRS, CNES, LAM Marseille, France \label{inst3}
\and
\centering    Université Paris-Saclay, Université Paris Cité, CEA, CNRS, AIM, F-91191 Gif-sur-Yvette, France\label{inst8}
\and
\centering Dipartimento di Fisica, Universit\`a di Roma Tor Vergata, Via della Ricerca Scientifica 1, I-00133 Roma, Italy \label{inst4}
\and
\centering Universität Heidelberg, Interdisziplinäres Zentrum für Wissenschaftliches Rechnen, Im Neuenheimer Feld 225, 69120 Heidelberg, Germany
 \label{inst5}
\and
Dipartimento di Fisica e Astronomia, Alma Mater Studiorum - Universit\`a di Bologna, Via P. Gobetti 92/3, I-40129 Bologna, Italy \label{inst6}
\and
INAF-Osservatorio Astrofisico di Arcetri, Largo E. Fermi 5, I-50125 Firenze, Italy \label{inst7}
}

\abstract
  % context heading (optional)
   {H{\sc\,ii} regions represent a key evolutionary stage of massive star formation, during which ionizing radiation reshapes the surrounding interstellar medium. 
   Continuum observations from the far-infrared to centimeter wavelengths trace thermal dust emission, free-free radiation from ionized gas, and anomalous microwave emission (AME) likely from rapidly rotating nano-grains. 
   Interpreting these tracers jointly is essential for connecting observed spectral energy distributions (SEDs) to the underlying physical structure of evolving star-forming regions.}
  % aims heading (mandatory)
   {We aim to quantify how thermal dust, free-free emission, and AME evolve during the expansion of an H{\sc\,ii} region and to determine how each tracer correlates with gas and electron column densities across spatial scales.}
  % methods heading (mandatory)
   {We post-process radiation-magnetohydrodynamical simulations of a $10^{4}\,M_\odot$ molecular cloud performed with \textsc{RAMSES} using the Monte Carlo radiative transfer code \textsc{POLARIS}. 
   The extended framework self-consistently includes grain charging, rotational excitation and damping, radiative torques, rotational disruption, thermal dust emission, free-free emission from hot plasma, and AME from nano-composite grains. 
   Synthetic maps and SEDs are computed for six evolutionary snapshots at frequencies  $1$-$10^{4}\ \mathrm{GHz}$ and analyzed with multi-scale correlation diagnostics of emitted radiation relative to total gas and electron column densities.}
  % results heading (mandatory)
   {The morphology of the ionized emission evolve from a compact, optically thick H{\sc\,ii} region to an extended, optically thin configuration. Free-free emission transitions at about $15\ \mathrm{GHz}$ from a $\nu^{2}$ to a $\nu^{-0.1}$ spectral regime depending on optical depth effects and increasingly traces the electron column density as the region expands in agreement with observations of H,\textsc{ii}  regions. The peak of the thermal dust emission shifts from mid-infrared toward the far-infrared as colder material contributes to the integrated SED, while maintaining a strong correlation with total gas column density. AME exhibits peak frequencies between $\sim26$ and $\sim37$\,GHz depending on evolutionary stage and environment, with scale-dependent correlations to both gas and electron distributions. Rotational disruption is identified as the primary dust-destruction mechanism influencing both mid-infrared dust emission and the high-frequency tail of the AME.}
  % conclusions heading (optional)
   {Our results demonstrate that thermal dust, free-free emission, and AME  provide complementary and scale-dependent diagnostics of stellar feedback in evolving H{\sc\,ii} regions. Self-consistent multi-frequency modelling is therefore crucial for interpreting current and future continuum surveys of star-forming clouds.}
  \maketitle

%________
\section{Introduction}
\label{sect:introduction}
Despite decades of theoretical and observational progress, the process of star formation remains poorly understood. Fundamental questions persist concerning how molecular clouds fragment, how turbulence, magnetic fields, and radiation feedback regulate collapse, and how newly formed massive stars influence their parental environment \citep{VazquezSemadeni2025,Beuther2025}. In particular, the interplay between gravity, magneto–hydrodynamics, radiative heating, and stellar feedback determines whether star formation proceeds efficiently or is quenched \citep{Traficante2020, Peretto2023, Suin2025}, yet the relative importance of these processes is still debated \citep{KlessenGlover2016,Federrath2016, Luo2024}.

In the process of star formation, H{\sc\,ii} regions represent a crucial evolutionary stage and mark the onset of ionizing radiation from young massive stars and thus the transition from embedded star formation to strong radiative feedback. The expansion of H{\sc\,ii} regions reshapes the surrounding interstellar medium (ISM), compresses molecular gas, alters dust properties, and drives dynamical instabilities. Hence, observations of such regions provide a direct probe of feedback physics and its role in regulating star formation up to the parsec
scales \citep{Zhang2021}. To connect observations with theory, it is essential to interpret the emission from star-forming regions in terms of the underlying physical state of gas and dust.

Continuum observations over a broad wavelength range are particularly powerful in this context. Far-infrared (FIR) and sub-millimeter (sub-mm) emission primarily trace thermal radiation from interstellar dust grains heated by stellar radiation and the interstellar radiation field. For instance, typical \textit{Herschel} bands near $70\ \mu\mathrm{m}$ are
sensitive to the emission of relatively warm dust components associated with embedded sources and active star formation, while longer wavelengths such as $250\ \mu\mathrm{m}$ predominantly trace colder dust in molecular clouds and dense filaments \citep{Molinari2010, Motte2010}. Together, these bands constrain dust temperature, column density, and the distribution of the cold and dense star-forming material \citep{Arzoumanian2011, Traficante15, Elia2017, Elia2021}.  

At centimeter wavelengths, free–free emission from ionized gas provides complementary information. Optically thin free–free radiation traces the emission measure of H{\sc\,ii} regions and therefore the ionizing photon output of massive stars, offering a direct measure of ongoing high-mass star formation as free-free emission is less affected by dust extinction \citep{Yang2023, Patel2023}. The combination of thermal dust and  free–free emission thus links the cold molecular reservoir to the ionized feedback phase.

However, observations of star-forming regions may further contain a component coming from anomalous microwave emission (AME). AME is widely interpreted as electric dipole radiation from rapidly rotating nanoparticles (“spinning dust”) \citep{DraineLazarian1998,Hoang2016} and peaks typically in the frequency range $\sim 10$–$60$\,GHz. Observationally, AME is often correlated with infrared dust tracers, yet the AME strength depends on the abundance, size distribution, and charge state of the smallest grains, as well as on the local radiation field and gas conditions. While early models focused primarily on polycyclic aromatic hydrocarbons (PAHs) and small carbonaceous grains, recent theoretical and observational developments have motivated the consideration of silicate bearing nano-particles as alternative or dominant AME carriers.

A key argument arises from interstellar abundance constraints.
A substantial fraction of cosmic carbon is locked in CO or large carbonaceous grains, limiting the abundance of free-flying PAH-sized particles available to emit spinning dust radiation \citep[see e.g.][]{AliHaimoud2009,Hoang2016}. Silicate-forming elements (Si, Mg, Fe), however, are almost entirely depleted into dust in the diffuse interstellar medium \citep{WeingartnerDraine2001}. If a modest fraction of this material exists in nanometer-sized silicate particles, their cumulative emission can match AME intensities without violating abundance constraints \citep{Hoang2016,HensleyDraine2017}. Considering the larger intrinsic dipole moments and the greater resilience to energetic environments we focus on AME modeling merely invoking composite grains as the dominant emitters in H{\sc\,ii} regions. 
Consequently, AME can either act as an additional tracer of the small-grain population and the microphysical environment of star-forming regions, or as a foreground contaminant that complicates the interpretation of free–free emission and hence star-formation diagnostics \citep{Planck2014}.

A self-consistent interpretation of multi-wavelength continuum data therefore requires models that simultaneously treat dust emission, ionized gas emission, grain charging, and dust evolution processes. In this work we address these challenges using synthetic observations of a star-forming cloud complex evolving into an H{\sc\,ii} region \citep{Suin2024,Suin2025}. We employ snapshots from a numerical simulation performed with the code \textsc{RAMSES} \citep{Teyssier2002} and
post-process them with the radiative transfer code \textsc{POLARIS} \cite{Reissl2016,Reissl2019}. Here, we use an extended version of \textsc{POLARIS} that self-consistently includes grain charging, dust destruction, thermal dust emission, free–free emission, and anomalous microwave emission within a unified framework. This approach allows us to quantify how the different continuum components trace the physical evolution of star-forming regions and to assess the diagnostic and contaminating role of AME in the presence of ionized feedback.

This paper is structured as follows: In Sect.~\ref{sect:mhd_simulation}, we outline the properties of the numerical H\,{\sc ii} region simulation. The adopted dust model is introduced in Sect.~\ref{sect:dust_model}, followed by a description of the MC RT scheme in Sect.~\ref{sect:mc_scheme}. Grain charging, rotation, and destruction are discussed in Sects.~\ref{sect:grain_charge}, \ref{sect:grain_rotation}, and \ref{sect:dust_destruction}, respectively. The emission and radiative transfer scheme used to generate synthetic observations is described in Sect.~\ref{sect:synthetic_obs}. Finally, we discuss the caveats and results of our synthetic observations in Sects.~\ref{sect:caveats} and \ref{sect:results}, respectively.

%\begin{figure*}[]
%	\centering
%	\includegraphics[width=1.0\textwidth]{plots/dust_em_70.pdf}
%    \includegraphics[width=1.0\textwidth]{plots/dust_em_250.pdf}
%	\caption{ Synthetic dust emission maps at wavelengths of $\lambda = 70\ \mu\mathrm{m}$ (top row) and $\lambda = 250\ \mu\mathrm{m}$ (bottom row), shown for three different lines of sight (see Fig.~\ref{fig:column}). The H{\sc\,ii}  region is assumed to be located at a distance of $D = 1~\mathrm{kpc}$.  }
%	\label{fig:dust_em}
%\end{figure*}

%\begin{figure*}[]
%	\centering
%	\includegraphics[width=1.0\textwidth]{plots/free_em_1.3.pdf}
%    \includegraphics[width=1.0\textwidth]{plots/free_em_3.1.pdf}
%	\caption{ Same as Fig.~\ref{fig:dust_em}, but for free–free emission at frequencies $\nu = 1.3\ \mathrm{GHz}$ (top row) and $\nu = 3.1\ \mathrm{GHz}$ (bottom row).}
%	\label{fig:free_em}
%\end{figure*}
\begin{figure*}[]
	\centering
	\includegraphics[width=0.85\textwidth]{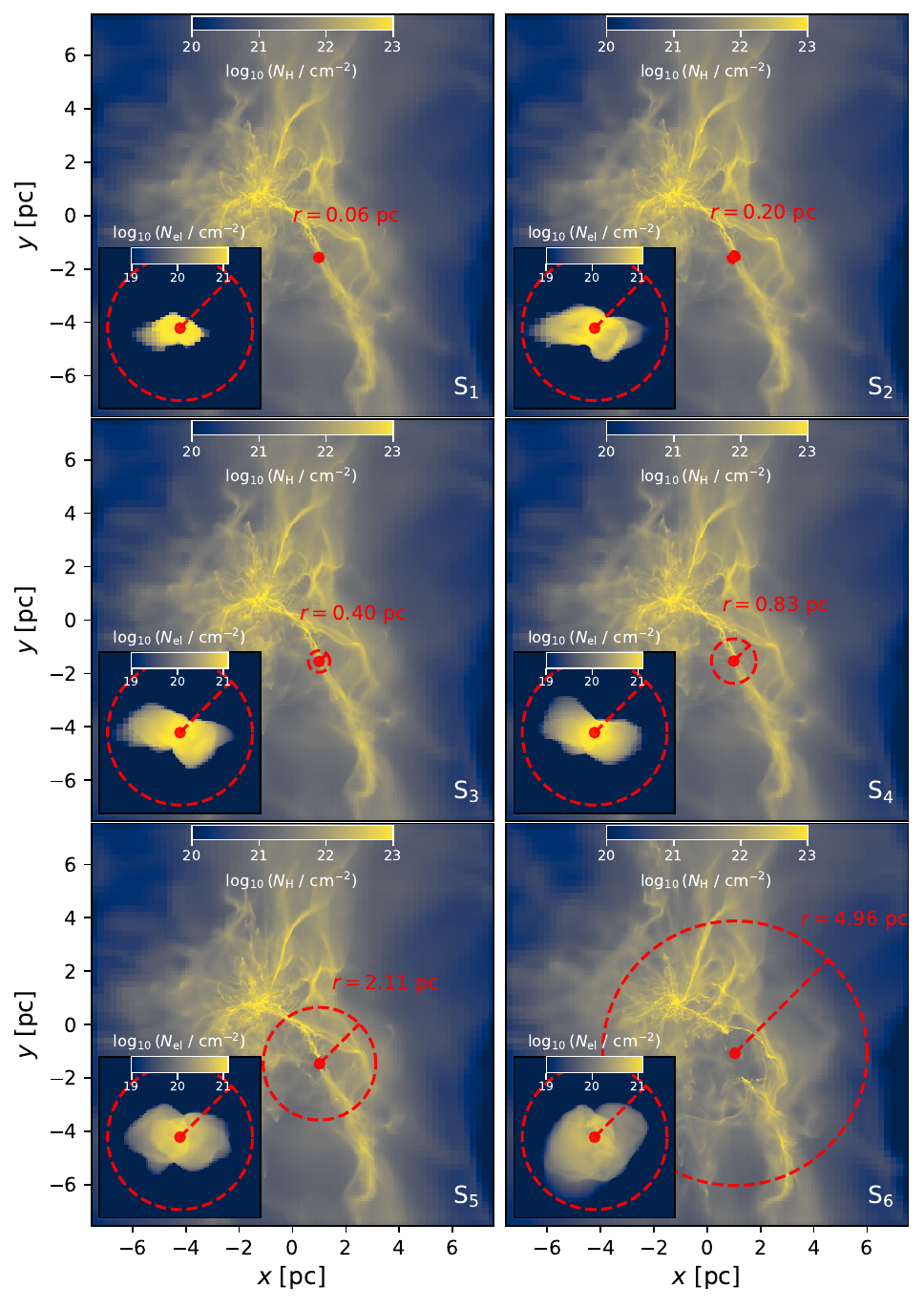}
	\caption{ Total gas column density $N_{\mathrm{H}}$ maps of the {\sc RAMSES} snapshots S$_1$ - S$_6$ representing the evolutionary stages of the H{\sc\,ii} region. Overplotted are the outer radius (red dashed circles) of the  spatial extension of the ionized gas and the red dot marks the position of the ionizing star within the simulation grid. The electron column density $N_{\mathrm{el}}$ is plotted in smaller panels on the characteristic scale of the H{\sc\,ii} region. We note that the red circles in the larger panels as well as the corresponding inset panels represent the same area.}
	\label{fig:column_density}
\end{figure*}

\section{RMHD simulation setup}
\label{sect:mhd_simulation}
We employed three-dimensional radiation-magnetohydrodynamical (RMHD)  simulations  of a $10^{4}\,M_{\odot}$ molecular cloud \citep{Suin2024,Suin2025,Suin2026} using the adaptive mesh refinement code {\sc RAMSES} \citep{Teyssier2002}. The computational domain is a cubic box of side length $L=30.4\ \mathrm{pc}$, initially resolved with $128^3$ cells and refined adaptively when the local Jeans length becomes smaller than 20 times the cell size. As we are interested in the expansion of H{\sc\,ii} regions, there is also a second flag for refinement when the ionization fraction jump is larger than a factor of ten between adjacent cells.

This setup allows for five refinement levels and a maximum spatial resolution of $\simeq 2 \times10^{-3}\ \mathrm{pc}$. Sink particles \citep{Federrath2010,Bleuler2014} are introduced at densities above $10^{8}$~cm$^{-3}$, enabling the simulations to follow gas accretion as well as stellar feedback.

The initial cloud was initially modeled as a Bonnor–Ebert  sphere of diameter $15.24\ \mathrm{pc}$, embedded in a lower-density medium. Its radial gas number density profile followed ${ n_{\mathrm{g}}(r)=n_{0}/[1+(r/r_{0})^{2}] }$, with $n_{0}=800$~cm$^{-3}$ and $r_{\mathrm{0}}=2.54\ \mathrm{pc}$. Thermal equilibrium was imposed through a parametrized cooling function \citep{Audit2005}, resulting in an initial temperature of $\sim 10\ \mathrm{K}$ in the densest regions. Turbulence was seeded with a Kolmogorov power spectrum and normalized such that the turbulent crossing time matched the central free-fall time. This corresponded to a Mach number of $\sim 6.7$, a velocity dispersion of $\sim 1.6\ \mathrm{km\ s}^{-1}$, and a virial parameter $\alpha_{\mathrm{vir}}\approx0.75$, consistent with observed star-forming clouds \citep{Rygl2013, Traficante2017}. The magnetic field was uniform and aligned with the $x$-axis, characterized by a mass-to-flux ratio $\mu = 2$. 

For the thermal and chemical evolution of the neutral gas the parametrization presented by \cite{Audit2005} is used. The ionization of the gas as well as the propagation of the ionizing  radiation from cell to cell is implemented into {\sc RAMSES} according to the scheme presented in \cite{Rosdahl2013} and \cite{Rosdahl2015}, respectively. This scheme follows, the propagation of photon fluxes only the energy bin $13.6 - 24.6\ \mathrm{eV}$, photons and gas coupling, as well as the ionization state and temperature of the gas self-consistently. As shown in \cite{Suin2026}, utilizing further energy bins for the radiation does not change the dynamical evolution of the emerging H{\sc\,ii} region. For simplicity the gas temperature equals the electron temperature $T_{\mathrm{g}} = T_{\mathrm{el}}$, a good approximation for H{\sc\,ii} regions considering the characteristic timescales for electron and gas to reach equilibrium \citep[see e.g.][]{Spitzer1978,Osterbrock2006,Pilyugin2007,Pilyugin2009}. 

Stellar feedback was incorporated via protostellar jets, radiative feedback, and sink particles representing individual stars. Here, a third of the accreted material is re-ejected in bipolar outflows with a $20^{\circ}$ opening angle and velocity set to $24\ \%$ of the local escape speed for each sink that reached $0.15\ M_{\odot}$.   
Once the entire cloud produced a mass of  $780\ \mathrm{M}_\odot$  of sinks in total, exactly one ionizing star is formed. The pool of a $780\ \mathrm{M}_\odot$ mass limit is chosen such that one can expect an ionizing star with a mass of $20\ \mathrm{M}_\odot$ typical for H{\sc\,ii}  regions \citep[see e.g.][]{Schaerer1997,Schaerer1998} in the stellar population in accordance with the Chabrier initial mass function \citep{Chabrier2003}. Hence, this simulation setup is particularly suited to study the impact of ionizing feedback on star-formation on a collapsing clouds and filaments. 

We select snapshots of the {\sc RAMSES} simulation in certain time intervals to cover the typical phases of H{\sc\,ii} evolution \cite{Habing1979,Wood1989}. In particular, we cover the ultra compact phase with a radius $r<0.1\ \mathrm{pc}$ and a gas density of $n_{\mathrm{g}} \approx 10^4 -  10^5\ \mathrm{cm}^{-3}$, the compact phase with 
$0.1\ \mathrm{pc} < r<0.5\ \mathrm{pc}$ and $n_{\mathrm{g}} \approx 10^3 -  10^4\ \mathrm{cm}^{-3}$, as well as the extended phase with $r>1\ \mathrm{pc}$ and  $n_{\mathrm{g}} < 10^3\ \mathrm{cm}^{-3}$. The hyper compact phase, characterized by $r<0.03\ \mathrm{pc}$, cannot be considered in this study due to the resolution limit of the {\sc 
RAMSES} AMR grid.

For convenience, we label the selected snapshots S$_{\mathrm{1}}$ - S$_{\mathrm{6}}$ in the following sections (see Table \ref{tab:MHD_param} for relevant parameters). Maps of gas column density $N_{\mathrm{H}}$ and electron column $N_{\mathrm{el}}$, respectively, for all of the distinct snapshots are shown in Fig.~\ref{fig:column_density}, where the radius $r$ of the H{\sc\,ii} region is defined by the spatial extension of free electrons within the simulation domain. 

For each snapshot the parameters of the total gas number density $n_{\mathrm{g}}$,  gas temperature and electron temperature, respectively, $T_{\mathrm{g}}=T_{\mathrm{el}}$, the number densities $n_{\mathrm{i}}$ of the ionized species, and the  stellar component represented by the sink particles, are utilized as input for the followup RT post-processing.

\section{Dust grain model}
\label{sect:dust_model}
\subsection{Optical properties}
Dust is not included in the {\sc RAMSES} RMHD simulation itself but is instead modeled in post-processing. The standard dust model for the ISM consists of two components, silicate and carbonaceous grains, which reproduce the observed absorption features well \citep{Mathis1977,DraineLi2001}. More recently, the {\sc Astrodust} model was introduced \citep{Draine2021ApJ,HensleyDraine2023}, in which multiple materials are incorporated into a single composite dust grain, providing an improved description of the observational constraints in the Milky Way. The corresponding optical properties are calculated from the refractive indices\footnote{The tabulated {\sc Astrodust} data are taken from \url{https://datacommons.princeton.edu/discovery/catalog/doi-10-34770-9ypp-dv78}.} using the code DDSCAT \citep{Draine1994,Draine2008,Flatau2012} for various grain radii $a$\footnote{Throughout this work, we denote the characteristic grain radius by $a$. While spherical grains are assumed for the emission calculations, deviations from spherical symmetry relevant to rotational dynamics are treated through parameterized prescriptions. Hence, for the calculation of grain rotation and charge offset, $a$ is identified with the effective radius $a_{\mathrm{eff}}$ of an equivalent-volume sphere, without explicitly resolving the grain shape.}.

For the optical properties per grain size and material the quantities of efficiencies of absorption $Q_{\mathrm{abs}}$ and scattering $Q_{\mathrm{sca}}$\footnote{We note that these quantities are dependent on grain size $a$ as well as wavelengths $\lambda$ although this dependency will not be explicitly labeled in the following paragraphs for simplicity.} as well as the parameter $g$ guiding the direction of scattered light are pre-calculated and tabulated logarithmically distributed over a wavelengths range in between the minimal wavelength at the Lyman limit of ${ \lambda_{\mathrm{min}} = 91.13\ \mathrm{nm} }$ and a maximal wavelength of ${ \lambda_{\mathrm{max}} = 2\ \mu\mathrm{m} }$, where dust extinction becomes negligible. 

\subsection{Grain size distribution}
In contrast to the diffuse ISM, dense cores with an onset in star-formation grain growth may already drive up to $\mu$m - sizes \citep{Valdivia2019}. Hence, we consider large grains (LG) between a minimal radius of $a_{\mathrm{min,LG}} = 10\ \mathrm{nm}$ and a maximal radius of $a_{\mathrm{max,LG}} = 1\ \mu\mathrm{m}$  following a power-law size distribution $\mathrm{d}n_{\mathrm{LG}} / \mathrm{d}a \propto a^{-3.5}$ \cite{Mathis1977}. In addition to LGs we also assume an ensemble of {\sc astrodust} nano-grains (NG) in between $a_{\mathrm{min,NG}} = 0.5\ \mathrm{nm}$ and a maximal radius of $a_{\mathrm{max,NG}} = 10\ \mathrm{nm}$ with a log-normal size distribution of 
\begin{equation}
    \frac{\mathrm{d}n_{\mathrm{NG}}}{\mathrm{d}a} \propto \frac{1}{a} \exp\left( - \frac{1}{2}\frac{\ln^2\left( a / a_{\mathrm{ref}} \right)}{\sigma^2}\right)
\end{equation}
with a typical reference radius of $a_{\mathrm{ref}} =0.35\ \mathrm{nm}$ and a variance of $\sigma = 0.4$ \citep{LiDraine2001,Hoang2016,HensleyDraine2023}. The full set of all dust parameters is listed in Table~\ref{tab:dust_param}.

Various growth and destruction mechanisms may considerably redistribute the grain sizes locally  \citep[see e.g.][for a comprehensive treatment of grain growth]{Kirchschlager2019,Hoang2020Galaxie,Calura2025}. However, due to the lack of the full history of the gas-dust feedback processes in between snapshots required to model such mechanisms accurately, the total dust distribution ${ \mathrm{d} n / \mathrm{d}a = \mathrm{d}n_{\mathrm{NG}}/\mathrm{d} a + \mathrm{d} n_{\mathrm{LG}}/\mathrm{d}a }$ is initially normalized for each snapshot such that we guarantee a typical dust mass to gas mass ratio of $1\ \%$ \citep{Bohlin1978,Draine2011} in each grid cell. Consequently, the total dust number density is defined by evaluating $n_{\mathrm{d}} = \int \mathrm{d} n / \mathrm{d} a \ \mathrm{d}a$.

\subsection{Electric dipole moment}
\label{sect:dipole_moment}
Certain grain materials as well as charged grains are associated with an effective electric dipole moment 
\begin{equation}
\mu_{\mathrm{dip}}^2 = \mu_{\mathrm{int}}^2 + \mu_{\mathrm{ch}}^2\,.
\end{equation}
Here, $\mu_{\mathrm int}^2 = N_{\mathrm{at}}\beta_{\mathrm{AD}}^2$ represents the intrinsic contribution arising from the material properties, where $N_{\mathrm{at}}$ is the average number of atoms in the grain (see Eq.~\ref{eq:number_atoms}). For the material parameter $\beta_{\mathrm{AD}}$, we adopt an effective value that accounts for the multiple material components of {\sc Astrodust} (see Appendix~\ref{app:dipole_ad}). The dipole moment ${ \mu_{\mathrm{ch}} = \epsilon |Z|e\,a }$ is a charge-induced contribution where $Z$ is the total charge on the grain surface and $e$ is the elementary charge. 
So far we assume that dust grains are perfectly spherical. However, stochastic processes lead to grain growth with surface irregularities and non-spherical shapes that may result in an average offset of the net grain charge $Ze$ from the center of mass and is quantified by a dimensionless charge-offset parameter $0.01 \lesssim  \epsilon \lesssim 0.1$ \citep[see][]{Purcell1975,DraineLazarian1998}. In this work we apply a value of $\epsilon = 0.5$ while for highly irregular nano-particles larger offsets are possible \citep{Hoang2016ApJ, Hoang2019ApJ}.

\subsection{Mechanical strength}
The tensile strength $\mathcal{S}_{\mathrm{max}}$ is the maximum stress a material can withstand under tension before mechanical failure occurs. We note that the magnitude of $\mathcal{S}_{\mathrm{max}}$ remains a matter of debate, particularly for solid grains and fractal aggregates \citep{Hoang2019Nat,Hoang2020Galaxie,Reissl2024}. Numerical simulations suggest that the tensile strength of porous grains may vary by several orders of magnitude \citep[][]{Seizinger2013,Tatsuuma2019ApJ,Reissl2024}. In this work, we adopt an upper tensile strength of $\mathcal{S}_{\mathrm{max}} = 1.0\times 10^{8}\ \mathrm{N\,m}^{-2}$, based on ab initio calculations and laboratory measurements \citep[][]{Gouriet2019,Hoang2019Nat,Jongchan2022}. This value is used as the default throughout this work unless stated otherwise. The impact of adopting lower values of $\mathcal{S}_{\mathrm{max}}$ is explored in the following sections.

\section{MC radiation field}
\label{sect:mc_scheme}
In order to calculate the radiation field (and subsequent quantities required for synthetic emission maps) for each RMHD snapshot, we apply the well tested RT code {\sc POLARIS} \citep{Reissl2016,Brauer2017,Reissl2019,Kobus2025} utilized in numerous publications\footnote{\url{https://portia.astrophysik.uni-kiel.de/polaris/}}. 
{\sc POLARIS} has the advantage to solve the RT problem per snapshot on the native AMR grid of {\sc RAMSES}. The code employs a Monte Carlo (MC) scheme to calculate the local radiation field  based on the {\sc RAMSES}  parameters, the dust model, and a stellar component as sources of radiation as MC simulation inputs. 

For the stellar component we use mass, age, and accretion rate provided by the {\sc RAMSES} sink particles itself where we have to deal with about 1400 particles dependent on the snapshot. For the later RT post-processing we need the full spectral distribution of the radiation emitted from all the sink particles. To infer a effective temperature $T_{\mathrm{eff}}$ and stellar radius $R_\star$ for the non-ionizing sinks we utilize MIST \footnote{MIST evolutionary tracks are provided by \url{https://waps.cfa.harvard.edu/MIST/index.html} and evaluated via the software package {\sc isochrones} \url{https://isochrones.readthedocs.io/en/latest/}} evolutionary tracks \citep{Dotter2016,Choi2016}, assuming solar metallicity to ensure consistency with the stellar age and mass \citep{Ekstrom2012}, reproducing a spectral luminosity $L_\lambda$ via the Stefan–Boltzmann law. For the one central ionizing star we also include the hydrogen-ionizing photon emission rate in the band $13.6 - 24.6\ \mathrm{eV}$ of $Q=8.86\times10^{47}\ \mathrm{s^{-1}}$ into our calculations. Matching the given $Q$ for a near-ZAMS $20\,M_\odot$ star %at $\sim\!5\times10^{5}\,$yr 
yields values characteristic of a OB main-sequence star, with $T_{\mathrm{eff}} \approx 3.3\times10^{4}\ \mathrm{K}$ and 
$R_\star \approx 7.5\,R_\odot$ corresponding to a bolometric luminosity of $L \approx 6.0\times10^{4}\ L_\odot$. Because the energy range of $Q$ is limited, there is a mild degeneracy where smaller and hotter or cooler and larger pairs of $(T_{\mathrm{eff}},R_\star)$ can match the fixed $Q$ in the order a few thousand Kelvin and $\sim$20\% in radius but the values chosen above are well within the range implied by \citet{Martins2005}, and compatible with Geneva tracks \citep{Ekstrom2012}. In addition to the stellar component, we use the canonical parametrization of the interstellar radiation field \citep[ISRF]{Mathis1983} with a $G_{\mathrm{0}}=1$ to illuminate the selected {\sc RAMSES} snapshots from the outside by a typical diffuse galactic bath of radiation \citep{Reissl2020}.

Photon packages are emitted in the {\sc POLARIS} MC run from each source individually where the energy and wavelengths of each package is sampled by $L_\lambda$. Each package is then scattered, absorbed and re-emitted based on the local densities and dust parameters $Q_{\mathrm{abs}}$, $Q_{\mathrm{sca}}$, and $g$ until it leaves the simulation domain. Finally, the spectral  energy density $u_\lambda =  4 \pi / c \times \sum | \vec{J}_\lambda |$ as well as the anisotropy parameter ${\gamma_\lambda =  |\sum  \vec{J}_{\lambda} | / \sum |  \vec{J}_{\lambda} | \in [0,1]}$ 
is assigned to each grid cell given the vector quantity of the mean intensity $\vec{J}_\lambda =  (4\pi)^{-1} \int \vec{I}_\lambda \mathrm{d}\Omega
$ over the solid angle where the intensity $\vec{I}_\lambda$ is directly resulting from the {\sc POLARIS} MC run.

\begin{table}
\caption{Model parameters of the {\sc RAMSES} snapshots labeled S$_1$ - S$_6$.}
\label{tab:MHD_param}
\centering
\begin{tabular}{|l|c|c|c|}
\hline
snapshot & age [Myr] & H{\sc\,ii} radius [pc] & phase \\
\hline
 S$_{\mathrm{1}}$ & 3.66 & 0.06 & ultra compact \\
 S$_{\mathrm{2}}$ & 3.67 & 0.20 & compact \\
 S$_{\mathrm{3}}$ & 3.68 & 0.40 & compact \\
 S$_{\mathrm{4}}$ & 3.69 & 0.83 & compact \\
 S$_{\mathrm{5}}$ & 3.67 & 2.11 & extended  \\
 S$_{\mathrm{6}}$ & 4.12 & 4.96 & extended \\
\hline
\end{tabular}
\end{table}

\begin{table}
\caption{Model parameters for dust grains.}
\label{tab:dust_param}
\centering
\begin{tabular}{|l|c|}
\hline
Parameter & value  \\
\hline
$a_{\mathrm{min,LG}}\ \mathrm{[nm]}$ & 10  \\
$a_{\mathrm{max,LG}}\ \mathrm{[nm]}$ & 1000  \\
$\rho_{\mathrm{gr}}\ \mathrm{[kg\ m^{-3}]}$                           & 3481  \\
$T_{\mathrm{sub}}\ \mathrm{[K]}$               & 1200  \\
$\eta$                           & -4  \\
$q$                              & -3.5  \\
$\mathcal{S}_{\mathrm{max}}\ \mathrm{[N\ m^{-2}]}$               & $1.0\times 10^{8}$  \\
$m_{\mathrm{0}}/m_{\mathrm{H}}$                          & 21.6  \\
$a_{\mathrm{min,NG}}\ \mathrm{[nm]}$ & 0.5  \\
$a_{\mathrm{max,NG}}\ \mathrm{[nm]}$ & 10   \\
$a_{\mathrm{ref}}\ \mathrm{[nm]}$  & 0.35   \\
$\sigma$                         & 0.4   \\
$\mathcal{W}\ \mathrm{[J]}$                & $1.17\times 10^{-18}$   \\
$\beta_{\mathrm{AD}}\ \mathrm{[C\ m]}$                          & $4.18\times 10^{-30}$   \\
$\epsilon $         & 0.5   \\
\hline
\end{tabular}
\end{table}

\section{Charged dust grains}
\label{sect:grain_charge}
The grain-charging prescriptions of \cite{WeingartnerDraine2001} provide material-dependent parameters for separate silicate and carbonaceous grains. To approximate the charging of composite grains within this framework, we treat {\sc Astrodust} as a simplified mixture consisting of $f_{\mathrm{sil}}=80.7\ \%$ silicate-oxygen material and $f_{\mathrm{C}}=19.3\ \%$ carbonaceous material by volume (see Appendix~\ref{app:dipole_ad}). The material-dependent charging parameters are calculated as volume-weighted averages of the corresponding silicate and carbonaceous values.

Considering our spherical dust grains of radius $a$ immersed in a plasma of temperature $T_{\mathrm{g}}$, electron number density $n_{\mathrm{e}}$, and ion number densities $n_{\mathrm{i}}$, and their corresponding masses $m_{\mathrm{e}}$ and $m_{\mathrm{i}}$, respectively, we calculate the local integer charge $Z\in\mathbb{Z}$ per grain size. Here, the charging is modeled as a charge-discharge process on the charge ladder. The steady-state probability distribution $f(Z)$ to find a particular grain with a charge $Z$ satisfies the master equation
\begin{equation}
f(Z) J_{\uparrow}(Z) = f(Z+1) J_{\downarrow}(Z+1)
\label{eq:master}
\end{equation}
with normalization
\begin{equation}
\sum_{Z=Z_{\mathrm{lower}}}^{Z_{\mathrm{upper}}} f(Z)=1
\label{eq:normalization}
\end{equation}
where $Z_{\mathrm{lower}}$ and $Z_{\mathrm{upper}}$ are the lower and upper
hard charge bounds, respectively (see also Sect. \ref{sect:coulomb_explosion}). 
 
The total upward and downward transition rates are
\begin{equation}
J_{\uparrow}(Z) = \sum_{i} J_{i}(Z) + J_{\mathrm{pe}}(Z)\,,
\end{equation}
and
\begin{equation}
J_{\downarrow}(Z) \;=\; J_{e}(Z) \;+\; J_{\mathrm{pdt}}(Z)\,,
\end{equation}
respectively, where $J_{\mathrm{i}}$ denotes the collisional current of ion species $\mathrm{i}$ to the grain, $J_{\mathrm{e}}$ the collisional electron current, $J_{\mathrm{pe}}$ the photoelectric emission rate of electrons from the grain, and $J_{\mathrm{pdt}}$ the photodetachment rate relevant only for the case of $Z<0$.

The collisional current for each Maxwellian distributed species $X\in\{i,e\}$ of mass $m_{\mathrm{X}}$, charge $q$ (e.g. ${q=+e}$ for singly charged ions and ${q=-e}$ for electrons), number density $n_{\mathrm{X}}$, and sticking probability $\mathcal{S}$ impinging on a grain of radius $a$ at gas temperature $T_{\mathrm{X}}$, is written \cite{DraineSutin1987} as follows
\begin{equation}
J_{\mathrm{X}}(Z) = n_{\mathrm{X}} S \pi a^{2} \sqrt{\frac{8 k_{\mathrm{B}}T_{\mathrm{X}}}{\pi m_{\mathrm{X}}}} \tilde{J}(\tau,\chi)\,.
\label{eq:Jcoll}
\end{equation}
where $\tilde{J}$ is the dimensionless enhancement factor with the reduced temperature $ \tau = 4\pi\varepsilon_{0} a k_{\mathrm{B}}T_{\mathrm{X}} q^{-2} $  and $\chi = Z e q^{-1}$ is the ratio of the charge on the grain to the charge on the colliding particle\footnote{We note that {\sc POLARIS} works internally in SI units and solves the RT equation in wavelengths and not frequency space. Hence, all parameters and physical quantities are presented in this paper accordingly.}. For the efficient evaluation of $\tilde{J}$ we use the piecewise parametrization as outlined in \cite{DraineSutin1987} capturing the attractive $\chi<0$ and the neutral limit $\chi=0$ as well as repulsive suppression $\chi>0$. For the sticking we follow 
\cite{DraineSutin1987} with a value close to unity for ions while smaller values are more likely for electrons due to grain curvature. In this work we utilize the fixed values of $S=1$ for ions and $S=0.5$ for electrons colliding with the grain surface.

The photoelectric emission rate is expressed using the MC spectral energy density $u_{\lambda}$ of the radiation field in combination with the dust absorption efficiency $Q_{\mathrm{abs}}$ and one obtains
\begin{equation}
J_{\mathrm{pe}}(Z) \;=\; \pi a^{2}\ \frac{1}{h}\int\limits_{\lambda_{\min}}^{\lambda_{\mathrm{pet}}(Z,a)} Y\times Q_{\mathrm{abs}}\, \lambda\,u_{\lambda}\ \mathrm{d}\lambda\,,
\label{eq:Jpe}
\end{equation}
where $h$ is the Planck constant $\lambda_{\mathrm{min}}$ is the shortest occurring wavelength in the MC run, $Y$ is the yield, and the photoelectric threshold is ${ \lambda_{\mathrm{pet}}(Z,a) = {hc}/E_{\mathrm{pet}} }$. The photodetachment rate is computed with the corresponding cross section $\sigma_{\mathrm{pdt}}$ as
\begin{equation}
J_{\mathrm{pdt}}(Z<0) = \frac{1}{h}\int\limits_{\lambda_{\min}}^{\lambda_{\mathrm{pdt}}} 
\sigma_{\mathrm{pdt}} \lambda\,u_{\lambda}\,d\lambda\,,
\label{eq:Jpdt}
\end{equation}
with the threshold $ { \lambda_{\mathrm{pdt}} = {hc}/E_{\mathrm{pdt}} }$ of excess attachment electrons. In practice, we use the commonly utilized parametrization for the photodetachment cross section $\sigma_{\mathrm{pdt}}$ as outlined by  \cite{WeingartnerDraine2001}.

The threshold energies $E_{\mathrm{pet}}$ and $E_{\mathrm{pdt}}$, respectively, are determined by the valence ionization potential $IP_{\mathrm{v}}$, the electron affinity $EA$, and an additional tunneling barrier $E_{\mathrm{min}}$ for sufficiently negative charge states. Defining the Coulomb energy 
${E_{\mathrm{C}} = e^2 /( 4\pi\varepsilon_{0}a)}$
and denoting the material work function as 
${ \mathcal{W} = f_{\mathrm{sil}}\mathcal{W}_{\mathrm{sil}} + f_{\mathrm{C}}\mathcal{W}_{\mathrm{C}} = 1.17\times 10^{-18}\ \mathrm{J} }$  (see Table \ref{tab:dust_param}), a convenient parameterization for $Z>0$ of ionization and affinity consistent with \cite{WeingartnerDraine2001} is
\begin{equation}
IP_{\mathrm{v}} = \mathcal{W} + \left(Z+\tfrac{1}{2}\right)\,E_{C}
\end{equation}
and
\begin{equation}
EA(Z,a) = \mathcal{W} + \left(Z-\tfrac{1}{2}\right)\,E_{C}\,,
\end{equation}
respectively. For $Z<0$ the valence ionization potential is taken as $IP_{\mathrm{v}} =EA(Z + 1,a)$. The additional barrier applied for $Z\le -2$ reads ${ E_{\mathrm{min}} = \theta\left(|Z+1|\right)\,E_{\mathrm{C}} \Phi(a,Z) }$, 
where ${\theta(x)\ =x/(1+\xi^{-1/2})}$ and $\Phi(a,Z)$ is a weak size-charge correction function, frequently implemented as ${\Phi(a,Z)=1-0.3\,(a/10\ \text{\AA})^{-0.45}|Z{+}1|^{-0.26}}$ truncated to $]0,1]$. Finally, the photoelectric and photodetachment thresholds are then evaluated via
\begin{equation}
E_{\mathrm{pet}} =
\begin{cases}
IP_{v} & Z\ge -1\,,\\[0.5ex]
IP_{v}+E_{\mathrm{min}}\,, & Z\le -2\,,
\end{cases}
\end{equation}
and
\begin{equation}
E_{\mathrm{pdt}} = EA(Z+1,a)+E_{\min}\, ,
\end{equation}
respectively. 

While the updated yield $Y$ presented in \citep{WeingartnerDraine2001} considers the ejection of  Auger electrons as well as electrons from the inner shells we argue that such effects should only be dominant for X-ray radiation where as for the UV band of the radiation field, the dominant term is merely the band contribution $Y=Y_{\mathrm{band}}(E,a,Z)$ modeled via
\begin{equation}
Y_{\mathrm{band}}(E,a,Z) =  y_{2}(E,a,Z)\ \min \{\ y_{0}(\Theta) \times y_{1}(a,E),\ 1\ \}\,,
\label{eq:Yband}
\end{equation}
where the functions are $\Theta = E-E_{\mathrm{pet}} + \Gamma(Z) E_{\mathrm{C}}$ and $\Gamma(Z)=Z{+}1$ for $Z\ge 0$ and $\Gamma(Z)=0$ for $Z<0$. For silicate components a convenient form for the bulk factor is
\begin{equation}
y_{\mathrm{0,sil}}(\Theta)=\frac{(1/2)(\Theta/\mathcal{W})}{1+5(\Theta/\mathcal{W})}\,,
\end{equation}
and for carbonacious ones
\begin{equation}
y_{\mathrm{0,C}}(\Theta)=\frac{9\times 10^{-3}(\Theta/\mathcal{W})}{1+3.7\times 10^{-2}(\Theta/\mathcal{W})}\,,
\end{equation}
that gives a total bulk factor of ${y_{0} = f_{\mathrm{sil}}y_{\mathrm{0,sil}} + f_{\mathrm{C}}y_{\mathrm{0,C}}}$ for {\sc Astrodust}.
The corresponding enhancement $y_{1}(a,E)$ that accounts for finite electron escape and photon attenuation lengths is expressed as
\begin{equation}
y_{1}(a,E)\;=\;\left(\frac{\beta}{\alpha}\right)^{2}\,
\frac{\alpha^{2}-2\alpha+2-2e^{-\alpha}}{\beta^{2}-2\beta+2-2e^{-\beta}}\,,
\end{equation}
where $\alpha= a(l_{\mathrm{e}}+l_{\mathrm{a}})/(l_{\mathrm{e}}l_{\mathrm{a}})$ and $\beta=a/l_{\mathrm{a}}$, Here, we use the values  $l_{\mathrm{e}}\approx 10\ \text{\AA}$ and $l_{\mathrm{a}}\approx 100\ \text{\AA}$ suggested by \citep{WeingartnerDraine2001}. The escape fraction $y_{2}$ of  electrons follows from the kinematic limits,
\begin{equation}
y_{2}(E,a,Z)\;=\;
\begin{cases}
1 &\mbox{if } Z<0\,,\\
\dfrac{E_{\mathrm{high}}^{2}\ \bigl(E_{\mathrm{high}}-3E_{\mathrm{low}}\bigr)}{\bigl(E_{\mathrm{high}}-E_{\mathrm{low}}\bigr)^{3}}&\mbox{otherwise}\,,
\end{cases}
\end{equation}
with an energy range in between
${E_{\mathrm{low}} = -\left(Z + 1 \right) E_{\mathrm{C}}}$ and 
${E_{\mathrm{high}} = E-E_{\mathrm{pet}}}$, respectively.

With these ingredients the steady-state distribution $f(Z)$ and ots normalization follows from Eq. \ref{eq:master} and \ref{eq:normalization}. Numerically, {\sc POLARIS} constructs $f(Z)$ on a dynamically expanding window $[Z_{\mathrm{lower}},Z_{\mathrm{upper}}]$ once the energy density $u_\lambda$ is known. Starting from a charge estimate near the balance point where $J_{\uparrow}\!\approx\!J_{\downarrow}$, the ratio above to all integers is propagated in a window using logarithmic values for numerical stability. The window is widened symmetrically until the edges are numerically negligible, quantified by the small tail criterion of $f(Z)<10^{-6}$ on the most outer bins and finally $f(Z)$ is normalized. %The final distribution \(f(Z)\) then yields the mean, mode, and dispersion of the grain charge used elsewhere in the model.

\section{Grain rotation}
\label{sect:grain_rotation}
Rotational damping and excitation of interstellar and circumstellar dust grains arise from multiple physical processes, including radiative torques collisions of charged particles, infrared (IR) photon emission, and photoelectric (PE) emission.  
Each process contributes to the  damping and excitation \citep[see][for further details]{DraineLazarian1998,AliHaimoud2009}. In this section we briefly present the environmental parameters controlling the damping and excitation rates of grain rotation.
 %The rotational dynamics of interstellar and circumstellar dust grains are governed by a balance between various damping and excitation mechanisms \citep{Draine1998ApJ508157}. In this section we summarize all these mechanisms involved in the spin-up process in HII regions. coefficients $F$ and $G$, which measure the relative torque efficiency compared to collisions with neutral hydrogen atoms 

\subsection{Radiative torques}
\label{sect_rat}
The interaction of LGs with the local radiation field may spin the grains most efficiently up to super-thermal rotation by radiative torques \citep[RATs,][]{LazarianHoang2007}. The RAT itself acting on the dust grain is given by \citep{Hoang2014}
\begin{equation}
\Gamma_{\mathrm{RAT}} =  \frac{1}{2} a^2 \int\limits_{\lambda_{\mathrm{min}}}^{\lambda_{\mathrm{max}}} \gamma_\lambda\, \lambda Q_{\mathrm{RAT}} u_\lambda \ \mathrm{d}\lambda\,,
\label{eq:JRAT}
\end{equation}
where $\gamma_\lambda$ is the spectra anisotropy factor, and $Q_{\mathrm{RAT}}$ is the RAT efficiency being evaluated by the canonical parametrization
\begin{equation}
 Q_{\mathrm{RAT}} \approx \begin{cases} 0.4   &\mbox{if } \lambda \leq 1.8\, a \\ 
		    0.4 \times \left(\frac{\lambda}{1.8\, a} \right)^{\eta}  & \mbox{otherwise} \end{cases}\, .
\end{equation}
We note that the exact value of the power-law exponent $\eta$ is not well constrained. Numerical calculations show that $\eta$  may be a material parameter  range of values comprised between ${\eta \in \{-2.6,-4\}}$  \citep{Hoang2014,Herranen2019,Jaeger2024}. For convenience we apply a typical exponent of $\eta = -3$ for all grain materials in this work.

By evaluating the characteristic drag timescale for compact spherical grains 
with a moment of inertia of $I_{\mathrm{gr}} = 8/15 \times \pi \rho_{\mathrm{gr}} a^5$ colliding with neutral gas as
\begin{equation}
\tau_{\mathrm{g}} = \frac{3}{4\sqrt{\pi}}\frac{I_{\mathrm{gr}}}{ m_{\mathrm{g}}\,n_{\mathrm{g}}\,v_{\mathrm{th}} \,a^4 }\,,
\end{equation}
one gets the total drag timescale
\begin{equation}
\tau_{\mathrm{d}} = \left(\tau_{\mathrm{g}}^{-1} + \tau_{\mathrm{IR}}^{-1}\right)^{-1} = \frac{\tau_{\mathrm{g}}}{1+F_{\mathrm{IR}}}\,,
\end{equation}
where $v_{\mathrm{th}}$ is the thermal velocity of the gas and
${ \tau_{\mathrm{IR}} = F_{\mathrm{IR}}\tau_{\mathrm{g}} }$ is the drag timescale due to emission of IR photons and $F_{\mathrm{IR}}$ is the dimensionless IR damping parameter (see Eq.~\ref{eq:FIR}). Consequentially, the maximal angular velocity $\omega_{\mathrm{rot}}$ LGs may rotate, follows then from the relation ${ \omega_{\mathrm{rot}}=\Gamma_{\mathrm{RAT}}\ \tau_{\mathrm{d}} / I_{\mathrm{gr}} }$. 

\subsection{Neutral-neutral grain-gas collisions}
\label{sect:FG_neutral}
In contrast to LGs, which are most efficiently spun up by RATs, NGs are more effectively accelerated to high angular velocities via interactions with gas particles. For NGs it is convenient to express the corresponding  coefficients of rotation as the dimensionless quantities $F$ for damping and $G$ for excitation, representing the ratio of the actual angular momentum loss and stochastic torque to the reference rate for collisions with neutral hydrogen atoms where we follow the procedure as outlined in \cite{DraineLazarian1998}.

For neutral gas–grain interactions, the dimensionless coefficients $F_{\mathrm{nn}}$ and $G_{\mathrm{nn}}$ are close to unity under typical interstellar conditions \citep[][]{DraineLazarian1998} and may be evaluated as the sum over all grain charges and neutral gas collision partner 
\begin{equation}
    F_{\mathrm{nn}} = \sum_{Z_{\mathrm{gr}}} f(Z_{\mathrm{gr}}) \sum_{\mathrm{n}}  \frac{n_{\mathrm{n}}}{n_{\mathrm{H}}} \left(\frac{m_{\mathrm{n}}}{m_{\mathrm{H}}}\right)^{1/2} \times \bigl[ \exp{ (- \epsilon_{\mathrm{n}}^2) } +\sqrt{\pi}\epsilon_{\mathrm{n}}\erf{(\epsilon_{\mathrm{n}}})\bigr]
\end{equation}
If the grain carries an electric charge $Z_{\mathrm{gr}}$, neutral particles experience an induced dipole potential characterized by the dimensionless parameter
\begin{equation}
    \epsilon_{\mathrm{n}}^2 = 
    \frac{\alpha_{\mathrm{n}} Z_{\mathrm{gr}}^{2} e^{2}}
         {2(4\pi\varepsilon_{0})^{2} a^{4} k_{\mathrm{B}} T_{\mathrm{g}}}\,,
\label{eq:epsilon_n}
\end{equation}
where $\alpha_{\mathrm{n}}$ is the gas polarizability (for the interested reader we refer to \citealt{Miller1978} as well as \citealt{Lide2009} for comprehensive list of polarizabilities of atoms and molecular species). 
%(e.g.\ $\alpha_{\mathrm H_2}\!\approx\!0.79\times10^{-30}$\,C\,m$^2$\,V$^{-1}$).  
For $\epsilon_{\mathrm{n}} \ll 1$ this correction is negligible whereas for charged grains it enhances both $F_{\mathrm{nn}}$ and $G_{\mathrm{nn}}$ moderately via attractive polarization forces.
\begin{equation}
\begin{split}
G_{\mathrm{nn}} = \sum_{Z_{\mathrm{gr}}} f(Z_{\mathrm{gr}}) \sum_{\mathrm{n}}  \frac{n_{\mathrm{n}}}{2 n_{\mathrm{H}}} \left(\frac{m_{\mathrm{n}}}{m_{\mathrm{H}}}\right)^{1/2}\times \qquad\qquad\qquad\qquad\qquad \\ \qquad\qquad
\left[ e^{-\epsilon_n^2} + 2\epsilon_n^2 + \frac{T_{\mathrm{d}}}{T_{\mathrm{g}}}
\frac{ e^{-\epsilon_n^2} + \pi^2 \epsilon_n \mathrm{erf}(\epsilon_n) }
     { e^{-\epsilon_e^2} + \pi^2 \epsilon_e \mathrm{erf}(\epsilon_e) }
\left( e^{-\epsilon_e^2} + 2\epsilon_e^2 \right)
\right]\,,
\end{split}
\end{equation}
where $\epsilon_e$ is similar to Eq.~\ref{eq:epsilon_n} but with the dust temperature $T_{\mathrm{d}}$ instead of the gas temperature $T_{\mathrm{g}}$. 
In the original model of \cite{DraineLazarian1998}  colliding gas particles are treated as being temporarily accommodated on the grain surface before desorbing and contributing to the net angular momentum of the NGs. For simplicity, we adopt the common approximation that the accommodation time is long compared to the thermalization timescale, so that impinging particles fully equilibrate with the grain surface and desorb with a Maxwellian velocity distribution characterized by the dust temperature $T_{\mathrm{d}}$, consistent with standard treatments of interstellar grain surface chemistry \citep{
HollenbachSalpeter1971,BurkeHollenbach1983,CazauxTielens2002,CuppenHerbst2005,Perets2007,CazauxSpaans2009,Reissl2023}.

\subsection{Neutral-ion grain-gas collisions}
\label{sect:FG}
In the case of neutral grains, i.e. $Z_{\mathrm{gr}}=0$, the grain rotation is merely modified by the electric field of the incoming ion. Here, we split the contribution to excitation by the part cause by incoming ions
\begin{equation}
G_{\mathrm{ni,in}} =f(Z_{\mathrm{gr}}=0) \sum_{\mathrm{i}}  \frac{n_{\mathrm{i}}}{n_{\mathrm{H}}} \left(\frac{m_{\mathrm{i}}}{m_{\mathrm{H}}}\right)^{1/2}\times
\left(1+\frac{3\pi^2}{4}\phi+\frac{1}{2}\phi^2\right),
\end{equation}
and the part of  evaporating ions 
\begin{equation}
G_{\mathrm{ni,ev}} =
\frac{T_{\mathrm{d}}}{2\,T_{\mathrm{g}}}\,F_{\mathrm{i}},
\end{equation}
where we define the dimensionless quantity
\begin{equation}
\phi^2 =
\frac{e^2}{4\pi\varepsilon_0}
\frac{2 Z_{\mathrm{i}}^{2}}{a k_{\mathrm{B}} T_{\mathrm{g}}}
\end{equation}
The total excitation coefficient is then
$G_{\mathrm{ni}} = G_{\mathrm{ni,in}} + G_{\mathrm{ni,ev}}$. the corresponding 
damping coefficient is given by
\begin{equation}
F_{\mathrm{ni}} =
f(Z_{\mathrm{gr}}=0) \sum_{\mathrm{i}}  \frac{n_{\mathrm{i}}}{n_{\mathrm{H}}} \left(\frac{m_{\mathrm{i}}}{m_{\mathrm{H}}}\right)^{1/2}\times 
\left(1+\frac{\pi^2}{2}\phi\right).
\end{equation}

\subsection{Charged-ion grain-gas collisions}
\label{sect:FG_ions}
For of grains with charge number $Z_{\mathrm{gr}}$ colliding with ions of charge $Z_{\mathrm{i}} e$, the Coulomb interaction modifies the impact rate through the dimensionless parameter
\begin{equation}
    \zeta = \frac{Z_{\mathrm{i}} Z_{\mathrm{gr}}\,e^{2}}              {4\pi\varepsilon_{\mathrm{0}}\, a k_{\mathrm{B}} T_{\mathrm{g}}}.
\end{equation}
where $\zeta < 0$ covers the attractive interaction between charged grain and ion while $\zeta > 0$ represents the repulsive one. The ion damping coefficient follow as
\begin{equation}
    F_{\mathrm{ci}} = \sum_{Z_{\mathrm{gr}}} f(Z_{\mathrm{gr}}) \sum_{\mathrm{i}}\frac{n_{\mathrm{i}}}{n_{\mathrm{H}}}\!\left(\frac{m_{\mathrm{i}}}{m_{\mathrm{H}}}\right)^{1/2}\! g_1(\zeta),
\end{equation}
and the excitation coefficients for incoming and evaporating ions are given by
\begin{equation}
    G_{_{\mathrm{ci,in}}} = \sum_{Z_{\mathrm{gr}}} f(Z_{\mathrm{gr}}) \sum_{\mathrm{i}}\frac{n_{\mathrm{i}}}{n_{\mathrm{H}}}\!\left(\frac{m_{\mathrm{i}}}{m_{\mathrm{H}}}\right)^{1/2}\! g_2(\zeta),
\end{equation}
and
\begin{equation}
G_{\mathrm{ci,ev}} =
F_{\mathrm{ci}}\,
\frac{T_{\mathrm{d}}}{2\,T_{\mathrm{g}}}\,
\frac{
\exp\!\left(-Z_{\mathrm{gr}}^{2}\epsilon_{\mathrm{i}}^{2}\right)
+ 2 Z_{\mathrm{gr}}^{2}\epsilon_{\mathrm{i}}^{2}
}{
\exp\!\left(-Z_{\mathrm{gr}}^{2}\epsilon_{\mathrm{i}}^{2}\right)
+ \sqrt{\pi}|Z_{\mathrm{gr}}|\ \epsilon_{\mathrm{i}}\,
\mathrm{erf}\!\left(|Z_{\mathrm{gr}}|\ \epsilon_{\mathrm{i}}\right)
}\,,
\end{equation}
respectively, where $\epsilon_{\mathrm{i}}$ is similarly defined as Eq.~\ref{eq:epsilon_n}. The focusing functions are defined by  $g_1(\zeta) = 1 - \zeta$ and $g_2(\zeta) = 1 -  \zeta + \zeta^2/2$ for the repulsive case with $\zeta\leq0$ and ${ g_1(\zeta)=g_2(\zeta)=\exp(-\zeta) }$ for the attractive case i.e. $\zeta>0$. These functions describe the enhancement or suppression of grain rotation upon the ion impact.

\subsection{Plasma drag}
\label{sect:FG_plasma}
Grains with a net dipole moment $\mu_{\mathrm{dip}}$ (see Sect.~\ref{sect:dipole_moment} for details) moving through an ionized medium experience long-range Coulomb interactions producing plasma drag and excitation, respectively where the corresponding excitation coefficient may be written as 
\begin{equation}
\begin{split}
G_{\mathrm{p}} =
 \frac{e^{2}}{4\pi\varepsilon_{0}} \sum_{\mathrm{i}}  \frac{n_{\mathrm{i}}}{n_{\mathrm{H}}} \left(\frac{m_{\mathrm{i}}}{m_{\mathrm{H}}}\right)^{1/2} 
\,
\frac{2 Z_{\mathrm{i}}^{2}}
{3 a^{4} (k_{\mathrm{B}} T_{\mathrm{g}})^{2}} \times \qquad\qquad\qquad \\ \qquad\qquad\qquad \sum_{Z_{\mathrm{gr}}} f(Z_{\mathrm{gr}})\, \mu_{\mathrm{dip}}^{2}
\left[
\ln\!\left(\frac{b_{\omega}}{a}\right)
+
\frac{1}{3}
\ln\!\left(
\frac{\min(b_q,\lambda_{\mathrm{D}})}{b_{\omega}}
\right)
\right].
\end{split}
\end{equation}
%Here, the parameters $b_{\omega} = v_{\mathrm{th}}/\omega_{\mathrm{th}}$ and $b_{q} = I v_{\mathrm{th}}/\hbar$ are 
Here, the parameters $b_{\omega}$ and $b_{\mathrm{q}}$ are the characteristic impact parameters for the short- and long-range interactions and 
the quantity 
${ \lambda_{\mathrm{D}} =
[
\varepsilon_{0} k_{\mathrm{B}} T_{\mathrm{g}}/(n_{\mathrm{el}} e^2)]^{1/2} }$
is the Debye length accounting for the shielding within the plasma. Following \cite{DraineLazarian1998}, the corresponding dimensionless damping term $F_{\mathrm{p}}$ is taken to be of similar magnitude as $G_{\mathrm{p}}$.

\subsection{Infrared photon emission recoil}
Emission of infrared photons by a rotating grain carries away angular momentum and produces both systematic damping and excitation. The corresponding dimensionless coefficients follow directly from energy–momentum conservation for radiative damping of dust grains \citep{DraineLazarian1998,AliHaimoud2009} via
\begin{equation}
F_{\mathrm{IR}} = 
\frac{3}{2\pi c^{2}}\,
\frac{1}{n_{\mathrm{g}} m_{\mathrm{g}} v_{\mathrm{th}} a^{2}}
\int\limits_{\lambda_{\min}}^{\lambda_{\mathrm{max}}}
\lambda^{2}\,
Q_{\mathrm{abs}} B_{\lambda}(T_{\mathrm{d}})\ \mathrm{d}\lambda, 
\label{eq:FIR}
\end{equation}
and
\begin{equation}
G_{\mathrm{IR}} = 
\frac{\hbar}{8\pi k_{\mathrm{B}} T_{\mathrm{g}}c}\,
\frac{1}{n_{\mathrm{g}} m_{\mathrm{g}} v_{\mathrm{th}} a^{2}}
\int\limits_{\lambda_{\min}}^{\lambda_{\mathrm{max}}}
\lambda 
Q_{\mathrm{abs}} B_{\lambda}(T_{\mathrm{d}})\ \mathrm{d}\lambda, 
\label{eq:GIR}
\end{equation}
where $B_{\lambda}(T_{\mathrm{d}})$ is the Planck function  modulated by the dust absorption efficiency $Q_{\mathrm{abs}}$.  

\subsection{Photoelectric emission recoil}
Photoelectric emission introduces additional damping and excitation torques through the recoil associated with ejected electrons. The gas heating per grain is computed following \citet{WeingartnerDraine2001,WeingartnerDraine2006}:
\begin{equation}
\Gamma_{\mathrm{pe}} =
\frac{\pi a^2}{h}
\int\limits_{\lambda_{\mathrm{min}}}^{\lambda_{\mathrm{max}}}
Q_{\mathrm{abs}} u_{\lambda} \lambda Y \langle E_{\mathrm{e}}(\lambda,Z_{\mathrm{gr}})\rangle\,
\mathrm{d}\lambda,
\label{eq:Gampe}
\end{equation}
where total photoelectron emission rate per grain $J_{\mathrm{pe}}$ is given by Eq~\ref{eq:Jpe}, $Y(\lambda,Z_{\mathrm{gr}})$ the photoelectric yield for a grain of charge $Z_{\mathrm{g}}$, and $\langle E_{\mathrm{e}}\rangle$ the mean kinetic energy of ejected electrons at infinity. The latter is approximated as
\begin{equation}
\langle E_{\mathrm{e}}(\lambda,Z_{\mathrm{g}})\rangle =
\max\!\left[0,\ \frac{h c}{\lambda}
- \mathcal{W} - \Delta E_{\mathrm{band}}
- e\Phi_{\mathrm{gr}}\right]
\end{equation}
where $\Delta E_{\mathrm{band}}$ a band-structure correction, and ${ \Phi_{\mathrm{gr}} = e(Z_{\mathrm{gr}}+1) / (4\pi\varepsilon_{0} a) }$ the grain electrostatic potential.

The corresponding dimensionless damping and excitation coefficients with the formulation of \cite{AliHaimoud2009}, are
\begin{equation}
F_{\mathrm{pe}} =
\frac{m_{\mathrm{e}}}{m_{\mathrm{g}}}
\frac{1}
{2\pi a^2 n_{\mathrm{g}}
\sqrt{2k_{\mathrm{B}}T_{\mathrm{g}}/(\pi m_{\mathrm{g}})}} \times \sum_{Z_{\mathrm{gr}}} f(Z_{\mathrm{gr}}) J_{\mathrm{pe}}
\end{equation}
and
\begin{equation}
G_{\mathrm{pe}} =
\frac{m_{\mathrm{e}}}
{4 n_{\mathrm{g}} a^2 k_{\mathrm{B}} T_{\mathrm{g}}
\sqrt{8\pi m_{\mathrm{g}} k_{\mathrm{B}} T_{\mathrm{g}}}} \times \sum_{Z_{\mathrm{gr}}} f(Z_{\mathrm{gr}})
\left(
\Gamma_{\mathrm{pe}}
+ e\Phi_{\mathrm{gr}}
J_{\mathrm{pe}}
\right)\,.
\label{eq:Gpe}
\end{equation}
The first term in Eq.~\ref{eq:Gpe} accounts for stochastic excitation by photoelectrons, while the second represents systematic work against the grain potential.

\subsection{Total rate coefficients and rotational temperature}
\label{sect:t_rot}
The processes driving grain rotation are additive \citep{DraineLazarian1998} and for NGs the total damping and excitation are sums of all contributions i.e.
\begin{equation}
    F = F_{\mathrm{nn}} + F_{\mathrm{ni}} + F_{\mathrm{ci}} + F_{\mathrm{p}} + F_{\mathrm{IR}} + F_{\mathrm{pe}}, 
\end{equation}
and
\begin{equation}   
    G = G_{\mathrm{n}} + G_{\mathrm{ni}} + G_{\mathrm{ci}} + G_{\mathrm{p}} + G_{\mathrm{IR}} + G_{\mathrm{pe}}.
\end{equation}
In thermal equilibrium the rotational temperature $T_{\mathrm{rot}}$, quantifying the mean rotational energy of the grain, is associated with its rms angular velocity $\langle\omega_{\mathrm{rot}}^2\rangle$  via ${ I_{\mathrm{gr}}\langle\omega_{\mathrm{rot}}^2\rangle = 3 k_{\mathrm{B}}T_{\mathrm{rot}} }$. However, including charge and interaction with the radiation field the $T_{\mathrm{rot}}$ may be evaluated by the gas temperature $T_{\mathrm{g}}$  and the coefficients $F$ and $G$ \citep{DraineLazarian1998} by 
\begin{equation}
    \frac{T_{\mathrm{rot}}}{T_{\mathrm{g}}}
    = \frac{2G}{F}
      \left[
        1+\sqrt{1+\tfrac{20}{3}\ \tfrac{F^{2}}{G}\ \tfrac{\tau_{\mathrm{g}}}{\tau_{\mathrm{ed}}}}
      \right]^{-1}.
\label{eq:t_rot}
\end{equation}
Here, the characteristic damping timescale due to dipole-emission is  
\begin{equation}
\tau_{\mathrm{ed}} = \frac{16}{5}\ \frac{\varepsilon_{0}\ \rho_{\mathrm{gr}}\,a^{5}\,c^{3}}
{\mu_{\mathrm{dip}}^{2}\ \omega_{\mathrm{rot}}^{2}},
\label{eq:tau_ed}
\end{equation}

In practice, $T_{\mathrm{rot}}$ is evaluated after the energy density $u_\lambda$ of the MC run is calculated and a grain charge is already assigned to each grain size per cell. A problem arises from the fact that $\tau_{\mathrm{ed}}$ depends itself on the angular velocity $\omega$ of the grains. Hence, an accurate solution of Eq.~\ref{eq:t_rot} may require multiple iterations \citep{AliHaimoud2009} but that would inevitably push the evaluation of $T_{\mathrm{rot}}$ per grain size and each cell beyond reasonable limits concerning runtime as the {\sc RAMSES} grids may contain up to $2.3\times 10^7$ individual cells. However, for most purposes one can assume that $\tau_{\mathrm{g}} \ll  \tau_{\mathrm{ed}}$ and subsequently ${ T_{\mathrm{rot}} / T_{\mathrm{g}} \approx F/G }$ \citep{DraineLazarian1998,AliHaimoud2009, Tram2020ApJ}. We note that this approximation might not apply in regions of the simulation domain with thin gas. In order to mitigate the situation, we estimate the ratio ${ \tau_{\mathrm g} / \tau_{\mathrm ed} \approx (3.5\times10^{-10}\ \mathrm{m}\,/\,a )^7\,/\,(n_{\mathrm g}\,/\, 10^{10}\ \mathrm{m}^{-3})}$ \citep{Hoang2010} if needed to guarantee a balance between runtime and precision. We note that for a few individual cells $T_{\mathrm{rot}}$ may still be off by a factor of $2-5$. Since peak wavelengths in the AME emissivity is ${ \lambda_{\mathrm{peak}} 
\propto T_{\mathrm{rot}}^{-1/2} }$  (see Sect.~\ref{sect:ame_emission}) we consider these cell's AME a neglectable contribution to the overall SED. 

Moreover, as a result of the strong radiation field within the H{\sc\,ii} region the dust and gas velocities may decouple and additional contributions to the coefficients $F$ and $G$ may arise \citep[see e.g.][]{Tram2020ApJ}. However, modeling such drift effects goes beyond the scope of this paper since we lack the complex history of gas-dust evolution in between the post-processed {\sc RAMSES} snapshots.

\section{Grain destruction processes}
\label{sect:dust_destruction}
In this section we briefly outline the considered mechanisms that may redistribute the original grain size distribution $\mathrm{d}n / \mathrm{d}a$.

\subsection{Coulomb explosion}
\label{sect:coulomb_explosion}
Hard  bounds are imposed to the range of allowed range of charges $Z$ avoid unphysical states. The lower bound is the smallest integer $Z$ for which the electron affinity is positive,
\begin{equation}
Z_{\min}\;=\;\min\{\,Z\in\mathbb{Z}\,; \,EA(Z,a)>0\ \}\,,
\label{eq:Zmin}
\end{equation}
i.e. $Z_{\min} = Z_{\mathrm{lower}}$ which ensures at least marginal binding of an excess electron to the grain. An upper bound is set by the Coulomb-explosion criterion using the maximal tensile strength $\mathcal{S}_{\max}$ of the grain material. Equating the electrostatic stress ${ p=(Z_{\mathrm{gr}}e)^{2}/\bigl(32\pi^{2}\varepsilon_{0}a^{4}\bigr) }$ to $\mathcal{S}_{\max}$ yields the floor
\begin{equation}
Z_{\max}\;=\;\left\lfloor\frac{4\sqrt{2}\ \pi\,a^{2}}{e}\ \sqrt{\varepsilon_{0}\ \mathcal{S}_{\max}}\ \right\rfloor\,,
\label{eq:Zmax}
\end{equation}
which provides a conservative material-dependent cap on the maximal attainable positive charge. 

For evaluating the effects of Coulomb explosion, we allow the upper bound ${ Z_{\mathrm{upper}}>Z_{\max}}$ in the charge distribution function $f(Z)$ but remove the fraction $\sum_{Z=Z_{\mathrm{min}}}^{Z_{\mathrm{max}}} f(Z) /\sum_{Z=Z_{\mathrm{min}}}^{Z_{\mathrm{upper}}} f(Z)$ from the dust and subsequently from the RT process as that fraction is destroyed by Coulomb explosion.

\subsection{Dust sublimation}
\label{sect:sublimation}
A grain undergoes sublimation if its temperature exceeds its sublimation temperature $T_{\mathrm{sub}}$  
\citep[see e.g.][]{Hoang2020Galaxie}. The sublimation temperature of carbonaceous materials is considerably higher than that of silicate components. Therefore, we adopt the lower sublimation temperature typical of silicate grains for {\sc Astrodust}, assuming that the entire grain may be destroyed once the silicate components begin to sublimate, even though some carbonaceous fragments may survive (see Table~\ref{tab:dust_param}).
Assuming thermal equilibrium between the absorbed and emitted energy, the dust temperature per cell can be calculated with \citep[see][for details]{Lucy1999,BjorkmanWood2001,Reissl2016} 
\begin{equation}
\begin{split}
\int\limits_{\lambda_{\min}}^{\lambda_{\mathrm{max}}}   \int\limits_{a_{\mathrm{min,NG}}}^{a_{\mathrm{max,LG}}}J_{{\lambda}} \frac{\mathrm{d} n}{\mathrm{d} a} Q_{\mathrm{abs}}\ \mathrm{d}a\ \mathrm{d}\lambda = \qquad\qquad\qquad\\ \qquad\qquad\qquad \int\limits_{\lambda_{\min}}^{\lambda_{\mathrm{max}}} \int\limits_{a_{\mathrm{min,NG}}}^{a_{\mathrm{max,LG}}} B_{\lambda}\left(T_{\mathrm{d}}\right)\frac{\mathrm{d} n}{\mathrm{d} a} Q_{\mathrm{abs}}\ \mathrm{d}a\ \mathrm{d}\lambda\,.    
\end{split}
\label{eq:Tdust}
\end{equation}
Consequently, we remove the dust component once the dust temperature $T_{\mathrm{d}}$ exceeds the characteristic sublimation temperature $T_{\mathrm{d}}>T_{\mathrm{sub}}$ (see also Table \ref{tab:dust_param}). %Dynamically 286 of 22956606 cells ...

\subsection{Rotational disruption}
\label{sect:RD}
A dust grain might become rotationally disrupted (RD) by centrifugal forces when the grain exceeds an critical angular velocity \citep{Hoang2019Nat,Hoang2020Galaxie,Reissl2024}  of
\begin{equation}
    \omega_{\mathrm{RD}} = \frac{2}{a}  \left(  \frac{\mathcal{S}_{\mathrm{max}}  }{ \rho_{\mathrm{gr}}  }  \right)^{1/2}\,.
\label{eq:omega_rot}
\end{equation}
This relation allows to determine the critical disruption radius $a_{\mathrm{RD}}$ numerically, considering RATs and the local radiation field. In our model, all LGs with radii $a > a_{\mathrm{RD}}$ are assumed to reach angular velocities sufficiently high to undergo RD. We therefore interpret $a_{\mathrm{RD}}$ as the upper size limit for disruption under the assumption of maximal rotational disruption. In reality, a fraction of LGs may become trapped at low-$J$ attractor points and thus remain at angular velocities below the disruption threshold \citep[see][for further details]{Hoang2025}. The corresponding size bins need therefore to be removed from the LGs and the remaining population is to be redistributed.

For simplicity, we follow the approach that the fragments of rotationally disrupted LGs again follow a power-law size distribution, extending down to the lower size range of the NGs. Given that the mass of a single spherical grain is $m(a)=\rho_{\mathrm{gr}}\frac{4}{3}\pi a^3$, mass conservation for the entire grain population requires
\begin{equation}
\int_{a_{\min,\mathrm{LG}}}^{a_{\max,\mathrm{LG}}}
a^{q+3}\,\mathrm{d}a
=
\int_{a_{\min,\mathrm{NG}}}^{a_{\mathrm{RD}}}
a^{\delta+3}\,\mathrm{d}a.
\label{eq:RD_mass_integrals}
\end{equation}
The new power-law index $\delta$ is then determined by numerically solving \citep[see][for further details]{Giang2020ApJ}
\begin{equation}
\frac{
a_{\max,\mathrm{LG}}^{q+4}
-
a_{\min,\mathrm{LG}}^{q+4}
}{q+4}
=
\frac{
a_{\mathrm{RD}}^{\delta+4}
-
a_{\min,\mathrm{NG}}^{\delta+4}
}{\delta+4}.
\label{eq:RD_delta}
\end{equation}
Consequently, all grains with radii $a>a_{\mathrm{RD}}$ are removed from the LG population with the initial power-law index $q$, while the remaining population is redistributed according to the new size distribution with power-law index $\delta$. Since the new distribution extends into the size range of the NGs, the overlapping contribution is added to the corresponding NG size bins, thereby accounting for the mass transfer from LGs to NGs. Moreover, in the case of RD of NGs, we assume that they fragment completely and that their constituent material is returned to the gas phase.

\subsection{Heating and sublimation of NGs}
\label{sect:subNG}
The stochastic heating and sublimation of NGs is dominated by the absorption of individual energetic photons from the local radiation field in the UV and optical range, $\lambda \!\sim\! 0.09- 1\ \mu{\mathrm{m}}$ \citep{DraineLi2001}. However, assigning an individual dust temperature to each bin of the NGs is not feasible in terms of memory. Hence, we estimate one critical radius $a_{\mathrm{crit}}$ per grid cell based on the local radiation field following the stochastic heating process as outlines e.g. in \cite{DraineLi2001} and \cite{Camps2015}. 

A photon of wavelength $\lambda$ carries an energy of ${ E_\gamma = h\,c/\lambda }$and the corresponding photon number density of the radiation field is determined from its 
spectral energy density $u_\lambda$ as $n_\lambda = u_\lambda\ \lambda/ (h\,c)$. The mean photon energy that is actually absorbed by a grain with an absorption cross
section of $C_{\mathrm{abs}} = \pi a^2 Q_{\mathrm{abs}}$ is therefore
\begin{equation}
\label{eq:meanE}
\langle E_\gamma\rangle =
\frac{\displaystyle \int_{\lambda_1}^{\lambda_2} C_{\mathrm{abs}}\,u_\lambda\ \mathrm{d}\lambda}
     {\displaystyle \int_{\lambda_1}^{\lambda_2} \dfrac{C_{\mathrm{abs}}\,u_\lambda\ \lambda}{h\,c}\ \mathrm{d}\lambda},
\end{equation}
which reduces to the field-weighted average
${ \langle E_\gamma\rangle\!\approx\!\int u_\lambda\ \mathrm{d}\lambda/
\int u_\lambda\ \lambda/(h\,c)\ \mathrm{d}\lambda }$ if $C_{\mathrm abs}$ varies only weakly
across the ultraviolet band. For the range of wavelengths we utilize $\lambda_1=\lambda_{\max}$, because photons with $\lambda < 91.13\ \mathrm{nm}$ are mostly absorbed by HI and HeI in the gas. The same for the upper limit $\lambda_2= 1\ \mu\mathrm{m}$, as longer wavelengths contribute merely to the equilibrium grain heating but not to stochastic spikes.

After the absorption event of one photon, the grain’s temperature jumps from its initial value $T_{\mathrm{i}}$ to the final value $T_{\mathrm{f}}$. This represents a change in internal energy of 
\begin{equation}
E_{\Delta\mathrm{T}}  = \int\limits_{T_{\mathrm{i}}}^{T_{\mathrm{f}}} C(T)\ \mathrm{d}T,
\end{equation}
where $C(T)$ is the heat capacity.  In the high-temperature regime ($T_{\mathrm{d}} \gg 100\ \mathrm{K}$) the Dulong–Petit limit applies and ${ C(T) \approx 3N_{\mathrm{at}} k_{\mathrm{B}} }$, with $N_{\mathrm{at}}$ being the average number of atoms
in the grain \citep{GuhathakurtaDraine1989}. This yields
\begin{equation}
T_{\mathrm{f}} \approx T_{\mathrm{i}} + \frac{E_\gamma}{3N_{\mathrm{at}} k_{\mathrm{B}}}\,.
\end{equation}
Expressing the average number of atoms within the grain as
\begin{equation}
N_{\mathrm{at}} = \frac{4 \pi a^3 \rho_{\mathrm{gr}}}{3 m_{\mathrm{0}}}\,, 
\label{eq:number_atoms}
\end{equation}
where $\rho_{\mathrm{gr}}$ is the material density and
$m_{\mathrm 0}$ is the mean atomic mass of the grain material  and by assuming that $T_{\mathrm{i}} \ll T_{\mathrm{f}}$ gives the condition
${ \langle E_\gamma\rangle > E_{\Delta\mathrm{T}} \approx 3 N k_{\mathrm{B}} T_{\mathrm{sub}} }$ and finally an approximation for a critical grain size
\begin{equation}
\label{eq:acrit}
a_{\mathrm crit} =
\left(
\frac{\langle E_\gamma\rangle\,m_{\mathrm{0}}}
     {4\pi\ \rho_{\mathrm{gr}}\,k_{\mathrm{B}}\,T_{\mathrm{sub}}}
\right)^{1/3}\,.
\end{equation} 
For NGs with and a typical UV photon energy
${ E_\gamma = 1.6\times10^{-18}\ \mathrm{J} }$,
Eq.~\ref{eq:acrit} yields a critical radius of $a_{\mathrm{crit}} \approx 0.6\ \mathrm{nm}$. All NGs with radius $a < a_{\mathrm{crit}}$ are assumed to be transiently heated above their sublimation temperature and removed from the RT process.

We emphasize that this approximation neglects the detailed temperature dependence of the specific heat and assumes that the absorbed photon energy is deposited as thermal energy throughout the lattice. However, this approximation provides a convenient sublimation estimate consistent with detailed numerical treatments \citep{Kirchschlager2019, Hoang2020Galaxie} and with stochastic-heating calculations of NGs temperature distributions \citep{DraineLi2001, Camps2015}.

\subsection{Thermal sputtering}
\label{sect:sputtering}
Thermal sputtering is one of the dominant destruction mechanisms for dust grains embedded in hot and ionized gas. In this process, energetic gas particles impact the grain surface and eject atoms from the solid, leading to a gradual reduction of the grain radius 
\citep{Tielens1994,Nozawa2006,Kirchschlager2019,Hoang2020Galaxie}.

For a spherical grain of radius $a$ and bulk density $\rho_{\mathrm gr}$, the rate of radius erosion due to thermal sputtering can be written as
\begin{equation}
\frac{\mathrm{d}a}{\mathrm{d}t}  = \frac{m_{\mathrm{0}} n_{\mathrm{g}} v_{\mathrm{th}} Y_{\mathrm{sp}} \left(T_{\mathrm{g}}\right)}
{4\ \rho_{\mathrm{gr}}}\,,
\label{eq:dadtsput}
\end{equation}
where $Y_{\mathrm{sp}}$ is the sputtering yield i.e. the number of atoms ejected per collision.

Since Eq.~\ref{eq:dadtsput} is independent of the grain radius, the erosion proceeds linearly in time. The characteristic thermal sputtering timescale is therefore defined as
\begin{equation}
\tau_{\mathrm{sp}} = \frac{a}{\lvert \mathrm{d}a/\mathrm{d}t\rvert} = \frac{4\ \rho_{\mathrm{gr}}\,a}
{m_{\mathrm{0}}\,n_{\mathrm{g}} v_{\mathrm{th}} Y_{\mathrm{sp}}\left(T_{\mathrm g}\right)}\,.
\label{eq:tausput}
\end{equation}
The temperature dependence of the sputtering yield is taken from the analytical fits presented by \citet{Kirchschlager2019} and \citet{Hoang2020Galaxie}, which are based on laboratory measurements and detailed sputtering calculations
\citep{Tielens1994,Nozawa2006}.
Here, we adopt the following parametrization
\begin{equation}
Y_{\mathrm{sp}}\left(T_{\mathrm g}\right)
=
Y_{\mathrm{max}}
\left[1 + \left(\frac{T_{\mathrm{c}}}{T_{\mathrm{g}}}\right)^{\gamma}\right]^{-1}
\left[1 - \exp\!\left(-\frac{T_{\mathrm{g}}}{T_{\mathrm{t}}}\right)\right],
\label{eq:yield}
\end{equation}
where $Y_{\mathrm{max}}=0.05$, $T_{\mathrm{c}}=1.5\times10^6\ \mathrm{K}$,
$T_{\mathrm{t}}=3\times10^5\ \mathrm{K}$ and $\gamma=2.5$.
This functional form ensures a sharp onset of sputtering above the threshold temperature and saturation of the yield at high temperatures and approximates the sputtering rates based on a mixture of silicate and carbonaceous grains (see Sect.~\ref{sect:grain_charge} for exact ratios) reported by \citet{Kirchschlager2019} to within $20\%$ over a gas temperature range of $10^5 - 10^7\ \mathrm{K}$, which is comparable to the intrinsic uncertainty of sputtering yield measurements \citep{Tielens1994,Nozawa2006}. Finally, we reduce the amount of grains in the size bin corresponding to the radius $a$ by factor of $\exp(-\Delta t / \tau_{\mathrm sp})$ where $\Delta t$ is the time difference in between the utilized {\sc RAMSES} snapshots (see Table~\ref{tab:MHD_param}).

\section{Synthetic observations}
\label{sect:synthetic_obs}
The {\sc RAMSES} snapshots provide the underlying physical parameters, such as
densities and temperatures. These are complemented by additional quantities
derived from the {\sc POLARIS} MC post-processing, including grain
charge, excitation, and destruction. Based on this combined information, we
generate synthetic multi-wavelength maps of free-free emission, thermal dust
emission, and AME. Here, {\sc POLARIS} solves the RT problem that read as follows.
\begin{equation}
\frac{\mathrm{d} I_\lambda}{\mathrm{d} \ell} = j_\lambda - \alpha_\lambda I_\lambda,    
\end{equation}
along distinct lines-of-sight (LOS). Here, $\mathrm{d} \ell$ is the infinitesimal path length, $I_\lambda$ is the intensity, and the RT coefficients $j_\lambda$ and $\alpha_\lambda$, respectively, are for emission and absorption (see the following sections). Numerically, {\sc POLARIS} utilizes a Runge–Kutta–Fehlberg (RFK45) solver with adaptive step-size \citep[see][for details]{Reissl2019}.

\subsection{Thermal dust emission}
\label{sect:dust_emission}
We model continuum radiative transfer through a dusty medium assuming local thermodynamic equilibrium (LTE) \ Hence, thermal dust emissivity follows from Kirchhoff's law as%AT{Question:is this approximation still good in a highly ionized medium? Maybe not...how complex would it be to extend this Equations to non-LTE regime?}.
\begin{equation}
    j_\lambda = \pi B_\lambda (T_{\mathrm{d}}) \int\limits_{a_{\mathrm{min,LG}}}^{a_{\mathrm{max,LG}}} a^2 \frac{\mathrm{d} n_{\mathrm{LG}}}{\mathrm{d} a} Q_{\mathrm{abs}}\, \mathrm{d} a\,.
\end{equation}
Equivalently, for a population of LGs the absorption and scattering are taken care by the coefficient 
\begin{equation}
    \alpha_\lambda = \pi \int\limits_{a_{\mathrm{min,SG}}}^{a_{\mathrm{max,LG}}} a^2 \frac{\mathrm{d} n_{\mathrm{LG}}}{\mathrm{d} a} \left(Q_{\mathrm{abs}} + Q_{\mathrm{sca}}\right) \mathrm{d} a\,.
\end{equation}
We emphasize the size distribution $\mathrm{d} n_{\mathrm{LG}}/ \mathrm{d} a$ may be modified by the grain destruction processes as outlined in Sect.~\ref{sect:dust_destruction}.

\subsection{Anomalous microwave emission}
\label{sect:ame_emission}
Environmental conditions also play a crucial role in determining whether
AME can arise. Small carbonaceous nano-particles,
such as PAHs, are highly susceptible to photodissociation and chemical
erosion in strong ultraviolet radiation fields, which are typical of
H,\textsc{ii} regions and photodissociation regions. As a result, the
population of ultrasmall grains responsible for spinning dust emission can
be significantly depleted in such environments. Silicate grains generally
possess higher binding energies and more robust lattice structures and are
therefore expected to survive longer than carbonaceous particles under
similar conditions \citep{Kirchschlager2019,Hoang2020Galaxie}. Nevertheless, even
silicate nano-grains can be destroyed by thermal sputtering in hot ionized
gas or be removed through coagulation and grain growth. In addition,
the strong ionization and enhanced electron densities within
H,\textsc{ii} regions modify the charging and rotational damping of
nano-particles, reducing the efficiency of electric dipole radiation.
Observational studies indeed find that AME is typically associated with
cold neutral or molecular gas and is weaker in strongly ionized regions
\citep{AME_Planck2016}. This combination of grain depletion and modified
excitation conditions likely explains why AME is prominent in molecular
and translucent clouds but often weak or absent in H,\textsc{ii}
regions.

Given the harsh conditions for PAHs even in the early stages of H\ \textsc{ii} evolution, we focus merely on {\sc Astrodust} grains as the carrier of AME in our RT post-processing consistent with recent modeling efforts \citep{HoangLazarian2016,HensleyDraineMeisner2016,HensleyDraine2017,Draine2021ApJ}. In detail, AME is produced by electric dipole radiation from rapidly rotating nanometer-sized dust grains \citep{DraineLazarian1998, Hoang2016,Hoang2019ApJ,Tram2020ApJ}. In this section we present the complete formulation of the AME emission as it is implemented in {\sc POLARIS}.

Within the limitations discussed in Sect.~\ref{sect:t_rot} for a thermalized grain rotation, the three-dimensional Maxwellian distribution \citep[see][for further details]{DraineLazarian1998,AliHaimoud2009} expressed as a function of wavelength is
\begin{equation}
f_{\mathrm{MW}}(\lambda, T_{\mathrm{rot}}) =
\frac{4}{\sqrt{\pi}}
\left(\frac{I}{2 k_{\mathrm{B}} T_{\mathrm{rot}}}\right)^{3/2}
\left(\frac{2\pi c}{\lambda}\right)^2
\exp\!\left[
-\frac{I}{2 k_{\mathrm{B}} T_{\mathrm{rot}}}
\left(\frac{2\pi c}{\lambda}\right)^2
\right]\,.
\label{eq:maxwell}
\end{equation} 
We note that the grains angular velocity is connected to the wavelength via  ${ \omega_{\mathrm{rot}} = 2\pi\,c/\lambda }$, Consequently, a part of all grains with a certain radius $a$ may reach the rotational disruption criteria defined in Eq.~\ref{eq:omega_rot} corresponding to a critical wavelengths ${ \lambda_{\mathrm{cr}} = \pi c\, a \times (\rho_{\mathrm{gr}}  / \mathcal{S}_{\max})^{1/2} }$. Hence, any AME at a  wavelengths $\lambda < \lambda_{\mathrm{cr}}$ may effectively be suppressed by means of rotational disruption by cutting off the corresponding tail of Eq.~\ref{eq:maxwell}.

The power radiated by a rotating electric dipole is
\begin{equation}
\langle P(\lambda) \rangle =
\frac{\mu_{\mathrm{dip}}^2}{18\pi\varepsilon_0 c^3}
\left(\frac{2\pi c}{\lambda}\right)^4\,.
\end{equation}
For simplicity, we assume that the rotating NGs do no have any preferential alignment direction and treat $\langle P(\lambda) \rangle$ as an orientation-averaged quantity \citep[for AME with aligned silicates we refer to][]{Hoang2016ApJ}. 

The corresponding emissivity per unit wavelength of a single dust grain is obtained by combining the dipole power with the rotational distribution function,
\begin{equation}
j_\lambda(a, Z) =
\frac{c}{4\pi\lambda^2}\,
\langle P(\lambda) \rangle\,
f_{\mathrm{MW}}(\lambda, T_{\mathrm{rot}}) .
\end{equation}
Subsequently, for a single grain with Maxwellian rotation, the emissivity $j_\lambda$ peaks at
\begin{equation}
\lambda_{\mathrm{peak}} =
2\pi c\,
\sqrt{\frac{I}{8 k_{\mathrm{B}} T_{\mathrm{rot}}}}
\propto a^{5/2}\,T_{\mathrm{rot}}^{-1/2}.
\end{equation}
This scaling explains why shifts to shorter wavelengths with increasing rotational temperature $T_{\mathrm{rot}}$. 

Considering, the full population of NGs with size distribution $\int \mathrm{d} n_{\mathrm SG}/\mathrm{d} a = n_{\mathrm d}$ and charge distribution $f(Z)$, the ensemble-averaged total  emissivity per hydrogen nucleus is
\begin{equation}
j_\lambda =
\sum_{Z_{\mathrm{gr}}}\int\limits_{a_{\min,SG}}^{a_{\max,SG}}
f(Z_{\mathrm{gr}})\,
j_\lambda(a, Z_{\mathrm{gr}})\,
\frac{\mathrm{d}n_{\mathrm SG}}{\mathrm{d}a}\, \mathrm{d}a .
\end{equation}
As $\lambda_{\mathrm{peak}} \gg \lambda_{\max}$ i.e. the peak wavelength of AME is much longer as the maximum wavelength $\lambda_{\max}$ where dust extinction is relevant, we perform the RT utilizing an absorption coefficient of $\alpha_\lambda=0$.

\subsection{Free-free emission}
\label{sect:free_free_emission}
Free-free emission is a result of the Coulomb acceleration of free electrons in the electric fields of ions. The change in specific intensity $I_\lambda$ of radiation as it propagates through an ionized medium if governed by the one-dimensional form of the radiative transfer assuming under LTE. The corresponding emissivity is given by \cite{Karzas1961} and \cite{Beckert2000} as
\begin{equation}
j_\lambda = \frac{16 \pi e^6}{3 m_{\mathrm{e}}^2 c^2} \left( \frac{2\pi}{3 k_{\mathrm{B}} m_{\mathrm{e}}} \right)^{1/2} Z_{\mathrm{i}}^2 n_{\mathrm{e}} n_{\mathrm{i}} T_{\mathrm{e}}^{-1/2} \frac{1}{\lambda^2}  \exp\left( - \frac{hc}{\lambda k T_{\mathrm{e}}} \right) \  \bar{g}_{\mathrm{ff}},
\end{equation}
where all parameters i.e. $n_{\mathrm{e}}$,  $n_{\mathrm{i}}$, $T_{\mathrm{e}}$, and $Z_{\mathrm{i}}$ are provided directly by the {\sc RAMSES} snapshots (see sect. \ref{sect:mhd_simulation}). 

The absorption coefficient for free-free processes, which quantifies the attenuation of radiation per unit length, is similarly given by \cite{Karzas1961,Beckert2000} 
\begin{equation}
\alpha_\lambda = \frac{8 \pi e^6}{3 h c^4 m_{\mathrm{e}}^2} \left( \frac{2\pi}{3 k_{\mathrm{B}} m_{\mathrm{e}}} \right)^{1/2} Z_{\mathrm{i}}^2 n_{\mathrm{e}} n_{\mathrm{i}} T_{\mathrm{e}}^{-1/2} \lambda^3 \left( 1 -  \exp\left( - \frac{hc}{\lambda k T_{\mathrm{e}}} \right) \right)\ \bar{g}_{\mathrm{ff}}.
\end{equation}
Here, the free-free Gaunt factor $\bar{g}_{\mathrm{ff}}(\lambda, T_{\mathrm{e}})$ provides a quantum mechanical correction to the classical Bremsstrahlung emissivity and absorption coefficients \citep[see e.g.][]{Karzas1961}. In this work, we adopt the thermally averaged Gaunt factor $\bar{g}_{\mathrm{ff}}$ as outlined in \cite{vanHoof2014}, which is derived from first-principles quantum mechanical calculations. Here, $\bar{g}_{\mathrm{ff}}$ depends on two key dimensionless parameters being the relative photon energy 
\begin{equation}
u = \frac{hc}{\lambda k_{\mathrm{B}} T_{\mathrm{e}}},    
\end{equation}
and the ratio of the ionic binding energy to the thermal energy, 
\begin{equation}
\gamma^2 = \frac{Z_{\mathrm{i}}^2 \, \mathrm{Ry}}{k_{\mathrm{B}} T_{\mathrm{e}}}\,.
\end{equation}

These two parameters, $u$ and $\gamma^2$, control the relative importance of quantum effects in the free-free process. In typical Galactic H{\sc\,ii} regions, where $T_{\mathrm{e}} \sim 8000 \, \mathrm{K}$ and hydrogen dominates the ionization balance ($Z_{\mathrm{i}}=1$), the parameter $\gamma^2$ is of order unity, while $u$ takes values $\ll 1$ at radio wavelengths (cm–mm) and approaches unity in the infrared to optical regime. 
\begin{figure*}[ht!]
	\centering
	\includegraphics[width=0.85\textwidth]{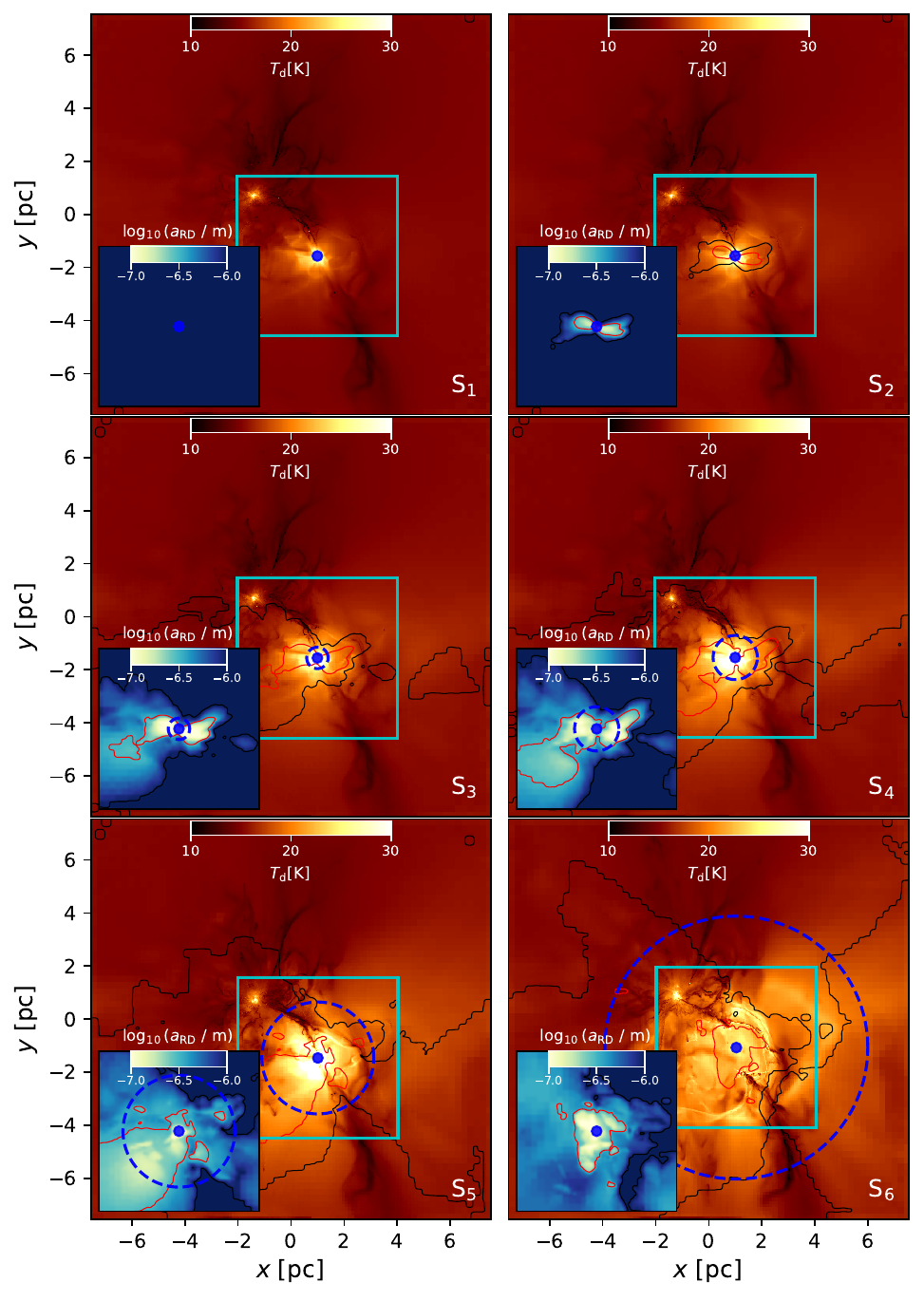}
	\caption{Same as Fig.~\ref{fig:column_density}, but showing the dust temperature $T_{\mathrm{d}}$ in the $xy$-plane containing the ionizing star (blue dot). Insets show the rotational disruption radius $a_{\mathrm{RD}}$. Red and black contours indicate $a_{\mathrm{RD}}=320\ \mathrm{nm}$ and $1\ \mu\mathrm{m}$, respectively, in both the large panels and the insets.}
	\label{fig:tdust}
\end{figure*}
% tbd note: star does not change in terms of stellar evolution but in the embedded phase S1 the radiation becomes efficiently re-distributed \cite{Reissl2018}
 \begin{figure*}[]
	\centering
	\includegraphics[width=0.85\textwidth]{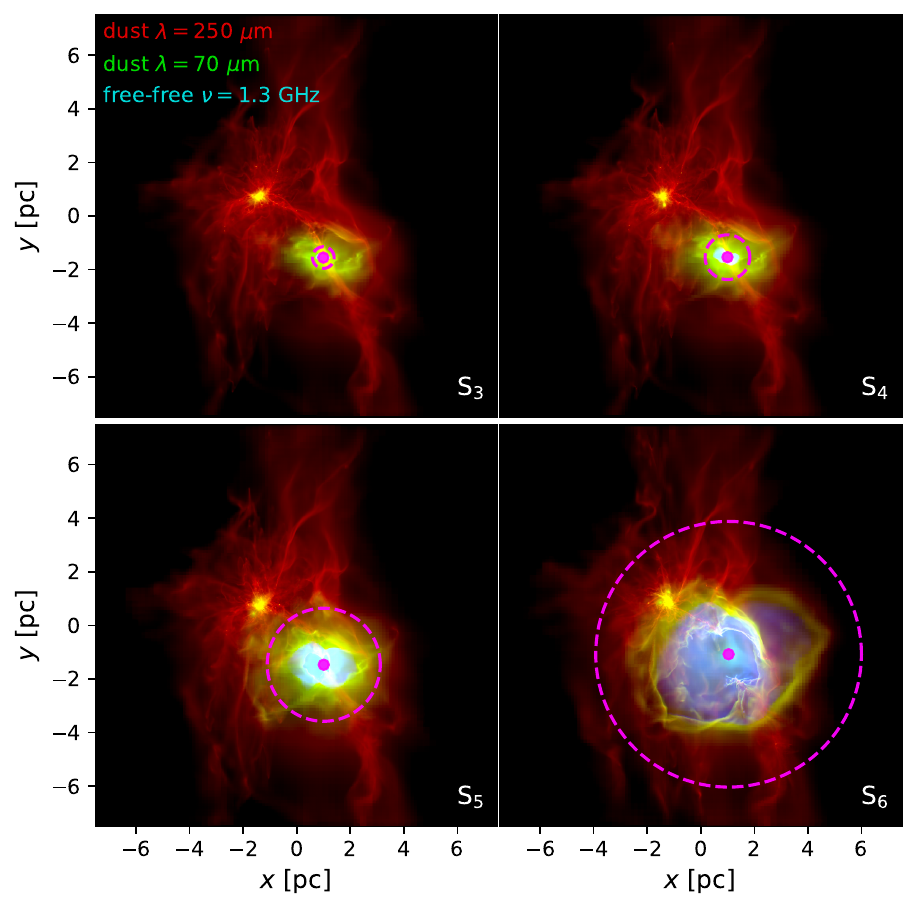}
	\caption{ Multicolor synthetic maps of the H{\sc\,ii} region for the snapshots S$_3$ - S$_6$ as seen along a distinct LOS, combining $250\ \mu\mathrm{m}$ (red) and $70\ \mu\mathrm{m}$ (green) thermal dust emission with $1.3\ \mathrm{GHz}$ free–free emission (blue). The position of the ionizing star and the extension of the ionized gas are marked by a purple dot and a dashed circle, respectively. The shown maps are along the same LOS and scale as in Fig.~\ref{fig:column_density}. }
	\label{fig:RGB_dust_free}
\end{figure*}

We emphasize that, in contrast to thermal dust emission and AME, the coefficients $j_\lambda$ and $\alpha_\lambda$ related to  free-free emission are independent of the radiation field and grain size distribution. 
Hence, $j_\lambda$ and $\alpha_\lambda$ can be calculated on the fly in the RT process itself. However, rather than evaluating the Gaunt factor analytically, we interpolate over the precomputed values provided by \cite{vanHoof2014}, which are tabulated\footnote{\url{https://data.nublado.org/gauntff/}} on a dense grid in $(Z_{\mathrm{i}}, u, \gamma^2)$. A three-dimensional interpolation routine is used to evaluate $\bar{g}_{\mathrm{ff}}$ for the local physical conditions in the AMR grid, ensuring both accuracy and computational efficiency in modeling free-free emission in the ionized gas within the {\sc RAMSES} snapshots.

\section{Caveats}
\label{sect:caveats}
The modelling framework presented in this work involves several
simplifications that should be considered when interpreting the results.

Dust destruction is treated using analytical prescriptions for thermal
sputtering and sublimation, including a parametrised temperature-dependent
yield. While this approach reproduces the correct order of magnitude and
temperature dependence \citep{Kirchschlager2019,Hoang2020Galaxie}, it neglects
variations in grain composition, surface structure, and non-thermal processes such as shock-induced sputtering. The treatment of stochastic heating and sublimation of nanograins relies on a single-photon approximation using an absorption-weighted mean photon energy. This neglects the full photon energy distribution and rare high-energy events, introducing uncertainties in the exact determination of critical grain sizes. 

Recently, \citet{Hoang2025} demonstrated that the efficiency of RD also depends on the fraction of grains  trapped at attractors associated with a low angular momentum that may remain below the critical angular velocity $\omega_{\mathrm{RD}}$. Consequently, the disruption radius $a_{\mathrm{RD}}$ should be interpreted as a threshold for possible disruption rather than a sharp boundary beyond which all grains are necessarily destroyed. However, accounting for the occupation of certain attractors requires a more complex treatment of grain alignment, magnetic relaxation, and stochastic rotational dynamics, which is beyond the scope of the present paper. Moreover, the adopted {\sc Astrodust} model does not uniquely constrain all the required grain properties, particularly the distribution of irregular grain shapes, which determines the RAT torque-component ratio, and the magnetic susceptibility, which governs the efficiency of magnetic relaxation.

Moreover, the size redistribution of rotationally disrupted grains remains poorly constrained. For dust shattering, it is commonly assumed that the resulting fragments follow a power-law size distribution \citep[see][and references therein]{Hirashita2009}. Hence, the transfer of mass from the large-grain end of the size distribution, where grains undergo rotational disruption, towards smaller size bins, and its impact on the resulting dust emissivity, may be modeled in a similar way by modifying the power-law index of the LG distribution \citep[][]{Giang2020ApJ,Hoang2020Galaxie}. However, the validity of this assumption requires further constraints from laboratory experiments, as N-body simulations indicate that the resulting fragments do not necessarily follow a power-law distribution but may instead efficiently break up into their smallest building blocks \citep[][]{Reissl2024}. Furthermore, it remains poorly constrained whether these building blocks can themselves undergo rotational disruption.

We emphasize that the physical origin of AME remains an active area of research and that modeling the rotational dynamics of NGs may involve additional complexities, including grain wobbling and quantum effects \citep[e.g.,][]{Hoang2010,DraineHensley2016}. Magnetic dipole emission from magnetic grains has also been proposed as an alternative or supplementary
contribution to AME \citep[e.g.][]{HoangLazarian2016}. In the present work, we adopt the original spinning-dust framework of \citet{DraineLazarian1998}, incorporating the refinements described by \citet{AliHaimoud2009} including the improvements in grain charging by \cite{WeingartnerDraine2001} and \cite{WeingartnerDraine2006}. This approach and variations thereof remain widely used in the literature
\citep[e.g.,][]{Tram2020ApJ} and provide a consistent basis for investigating the effects of dust evolution on AME. Although more detailed treatments may alter the predicted absolute emission levels and spectral shapes, we expect the principal trends identified here to remain qualitatively valid, as our analysis focuses on the relative changes produced by consistently modelling the dominant processes governing dust emission and destruction.

\section{Results}
\label{sect:results}

\begin{figure*}[]
	\centering
	\includegraphics[width=0.85\textwidth]{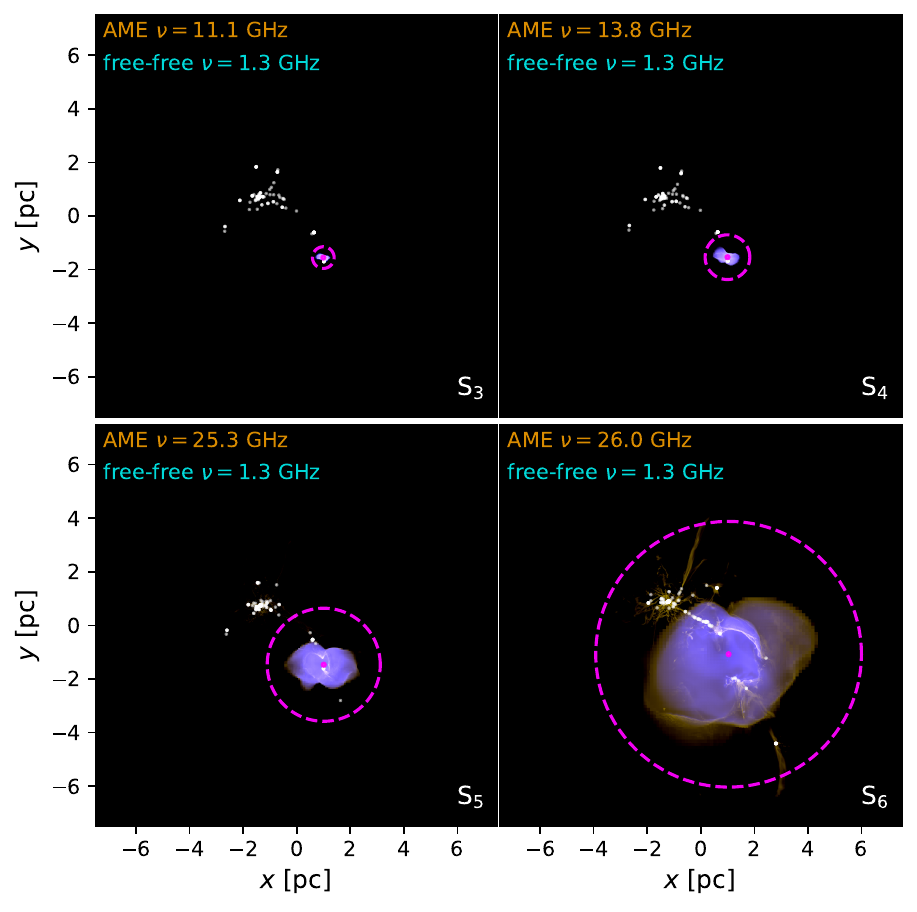}
	\caption{The same as Fig.~\ref{fig:RGB_dust_free} but for the AME (orange) at its peak frequency in comparison with the $1.3\ \mathrm{GHz}$ free–free emission (blue). The full stellar component in each snapshot is marked by white dots.  }
	\label{fig:RGB_AME_free}
\end{figure*}

%\begin{figure*}[]
%	\centering
%\includegraphics[width=1.0\textwidth]{old/SED_all_no_AME.pdf}
%	\caption{ SED (no AME)}
%	\label{fig:}
%\end{figure*}

\begin{figure}[]
	\centering
	\includegraphics[width=0.5\textwidth]{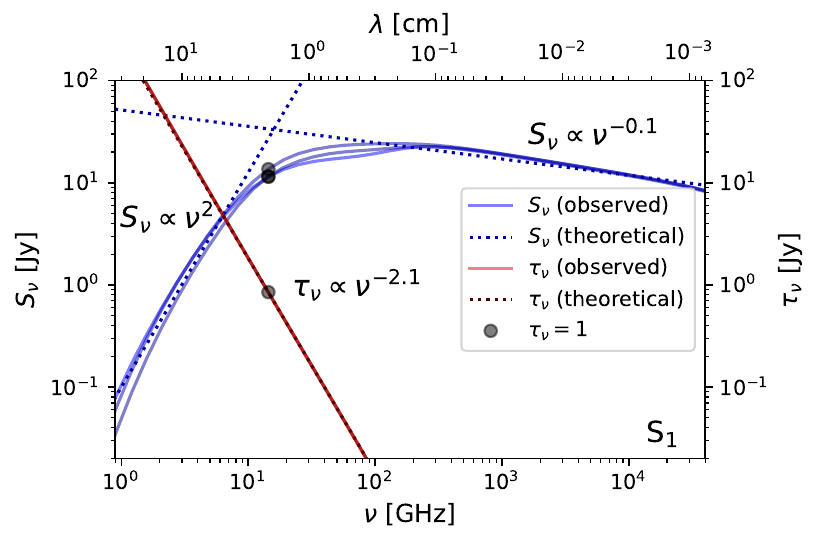}
	\caption{The SED of free-free emission $S_\nu$ (blue) and corresponding optical depth $\tau_ \nu$ of snapshot S$_1$ observed along three orthogonal LOSs and the (solid lines) in comparison with their theoretical predictions (dotted lines). }
	\label{fig:SED_free}
\end{figure}

\begin{figure*}[]
	\centering
	\includegraphics[width=1.0\textwidth]{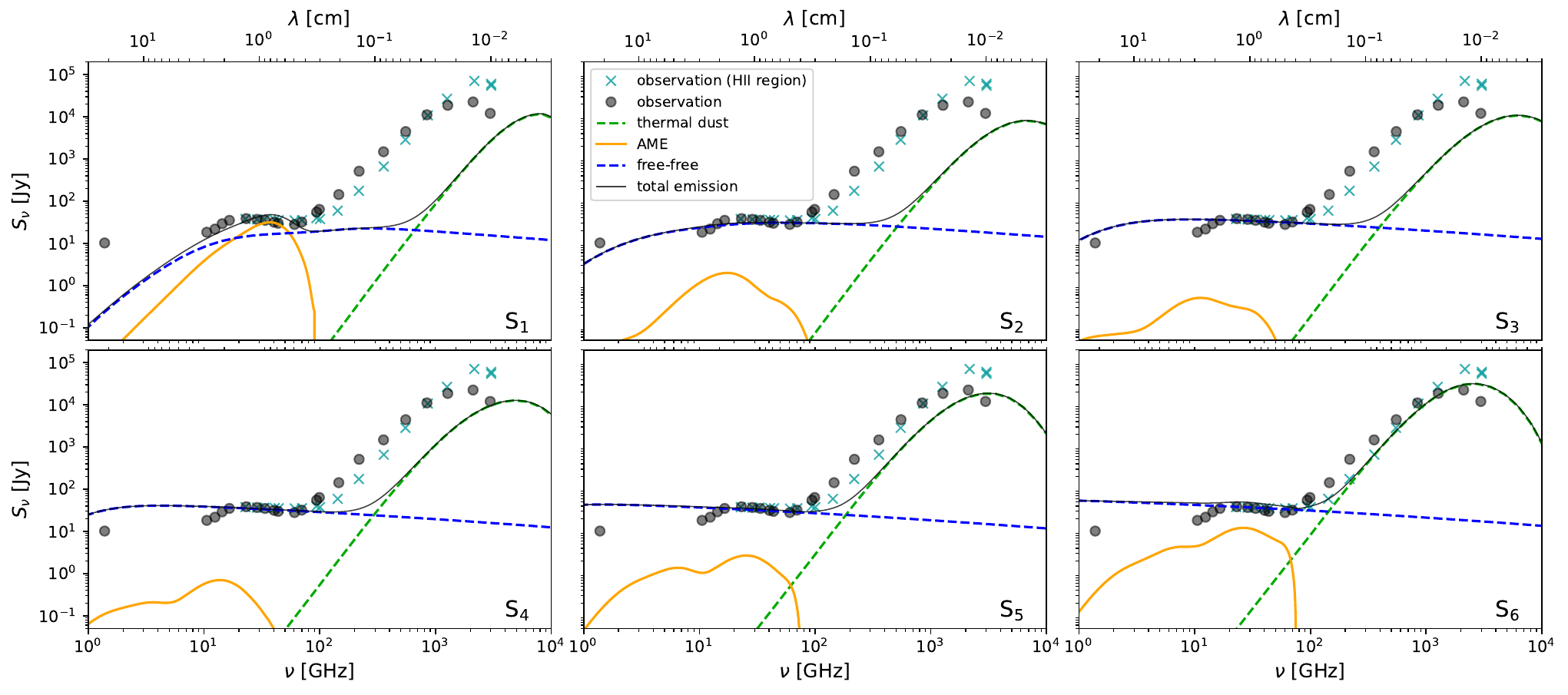}
	\caption{ SEDs over the frequency range $1\ \mathrm{GHz}$–$10^4\ \mathrm{GHz}$ of the free–free emission (blue dashed), the dust emission astrosilicate grains (green dashed), and the AME (orange solid) representing the evolutionary stages S$_1$ - $S_6$ of the simulated H{\sc\,ii} region. The SEDs are compared to observational data of a H{\sc\,ii} region in Orion (black dots) and a source in Perseus (cyan crosses), as presented by \cite{CollaborationPlanck2012}. All synthetic fluxes are derived within the characteristic radius of the ionized gas (see e.g. Fig.~\ref{fig:column_density}). We note that the fluxes of the actual observations are re-scaled to match the synthetic SEDs for a better comparison of characteristic features. }
	\label{fig:SEDs_aperture}
\end{figure*}

\begin{figure*}[]
	\centering
	\includegraphics[width=1.0\textwidth]{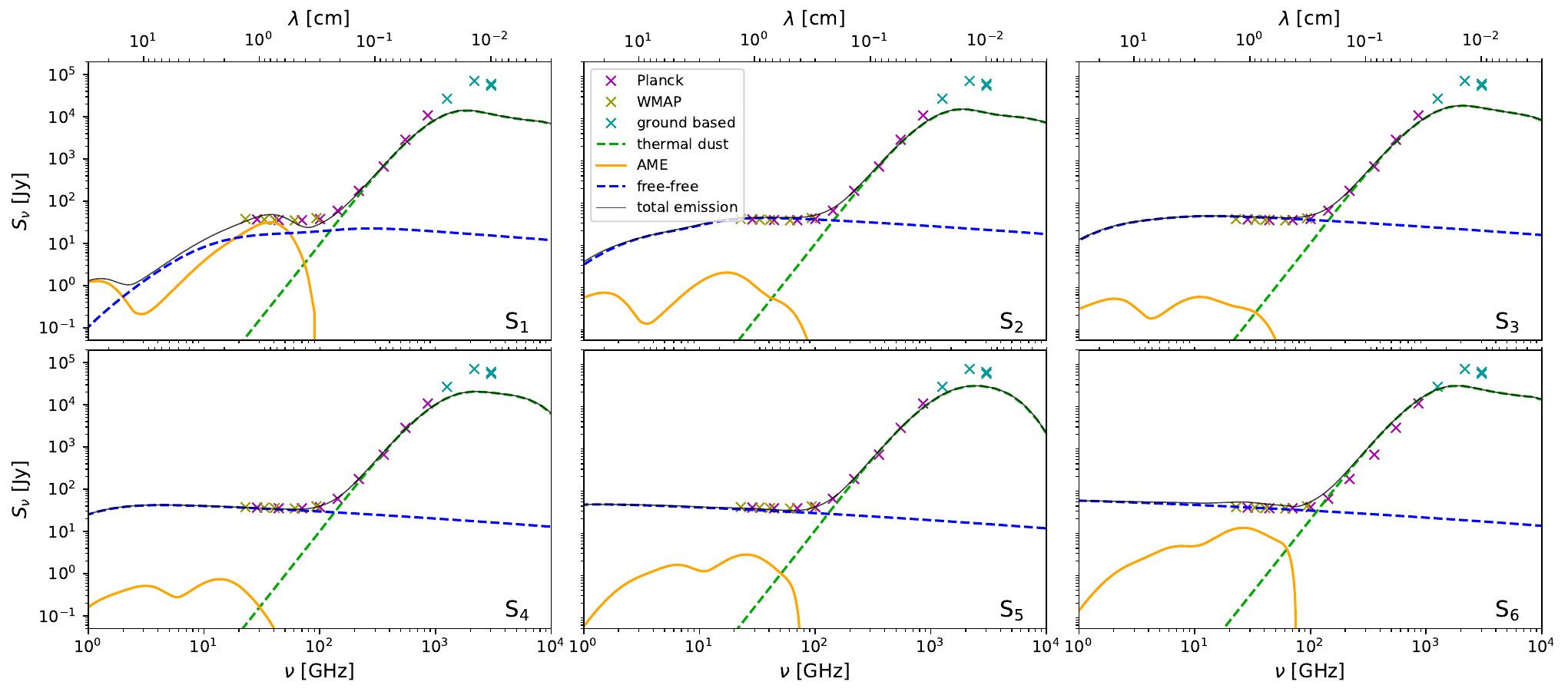}
	\caption{ The same as Fig.~\ref{fig:SEDs_aperture} but with all fluxes derived for the entire synthetic map. }
	\label{fig:SEDs_no_aperture}
\end{figure*}

\begin{figure*}[]
	\centering
	\includegraphics[width=1.0\textwidth]{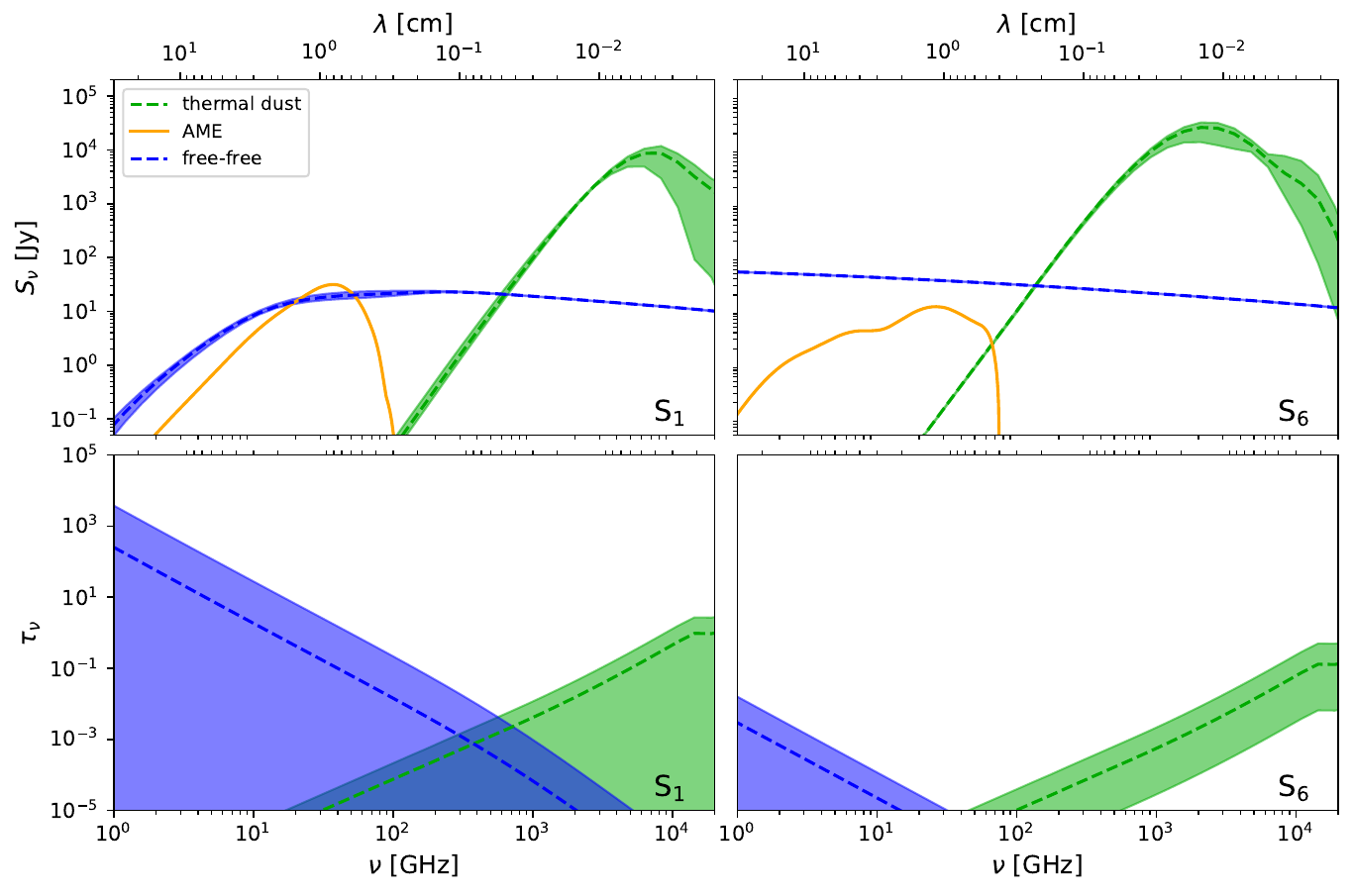}
	\caption{The average SEDs ($S_\nu$) of free–free emission (blue dashed), the dust emission {\sc astrodust} grains (green dashed), and the AME (orange solid), respectively, along three orthogonal LOSs for the distinct {\sc RAMSES} snapshots S$_1$ and S$_6$ (top row) in comparison with the corresponding optical depth ($\tau_\nu$, bottom row). Shaded areas represent the minimum and maximum values appearing in the emission maps.}
	\label{fig:SED_LOS}
\end{figure*}

\begin{figure*}[]
	\centering
	\includegraphics[width=1.0\textwidth]{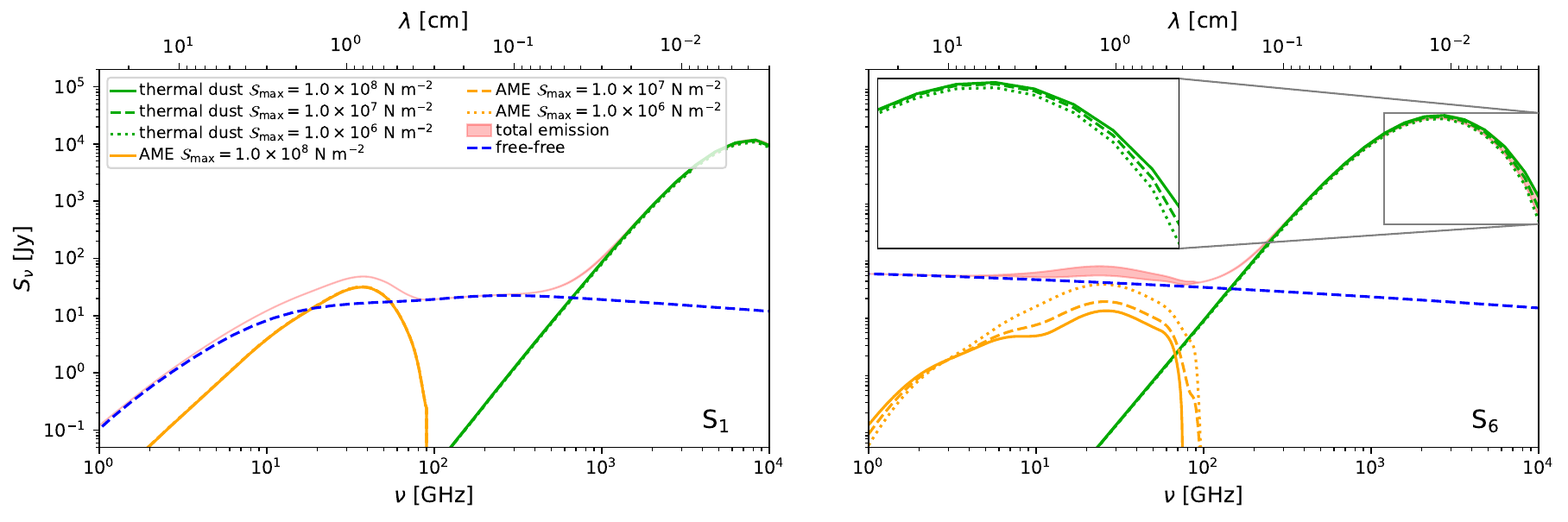}
    \caption{Spectral energy distributions (SEDs) of thermal emission from {\sc Astrodust} grains (green lines) and AME (orange lines) for the {\sc RAMSES} snapshots S$_1$ (left panel) and S$_6$ (right panel). The SEDs are compared for grain size distributions corresponding to tensile strengths of $\mathcal{S}_{\mathrm{max}}=1.0\times10^{8}\ \mathrm{N\,m}^{-2}$ (solid lines), $\mathcal{S}_{\mathrm{max}}=1.0\times10^{7}\ \mathrm{N\,m}^{-2}$ (dashed lines), and $\mathcal{S}_{\mathrm{max}}=1.0\times10^{6}\ \mathrm{N\,m}^{-2}$ (dotted lines). The red shaded region represents the range of total emission resulting from the different values of $\mathcal{S}_{\mathrm{max}}$. The blue dashed lines represent free-free emission. The inset in the right panel magnifies the thermal-emission peak to illustrate the marginal variation in the thermal emission of LGs. The panels should be compared with those shown in Fig.~\ref{fig:SEDs_aperture}.}
	\label{fig:dust_RD}
\end{figure*}

\subsection{Evolution of dust temperature and rotational disruption}

In Fig.~\ref{fig:tdust} we present the evolution of the dust temperature $T_{\mathrm{d}}$ and RD radius $a_{\mathrm{RD}}$ across snapshots S$_1$--S$_6$. We emphasize the the maximum of $T_{\mathrm{d}}$ may be much higher as $30\ \mathrm{K}$ resulting in dust sublimation around the ionizing star.

Although the intrinsic parameters of the ionizing star remain unchanged, the evolving density structure of the surrounding cloud modifies the local radiation field over time. In particular, stellar radiation is increasingly redistributed into the infrared as the H\,{\sc ii} region develops, changing both the spatial distribution of dust temperatures and the efficiency of RATs.

In the earliest snapshot, S$_1$, the ionizing star is still embedded in its dense natal cloud. Here, the most efficient re-distributions of stellar radiation \citep[see][]{Reissl2019} prevents efficient RD, and no significant disruption region is identified. As the H\,{\sc ii} region expands and the surrounding gas is dispersed, stellar radiation can propagate over larger distances and spin up dust grains through radiative torques. Consequently, RD becomes increasingly effective, as indicated by the expanding regions enclosed by the $a_{\mathrm{RD}}=320\ \mathrm{nm}$ and $1\ \mu\mathrm{m}$ contours. By S$_6$, RD affects a substantial fraction of the dust surrounding the ionizing star.

These results demonstrate that the efficiency of rotational disruption is governed not only by the intrinsic stellar luminosity but also by the evolving distribution of gas and dust, which determines the local radiation field and its interaction with the grains.

\subsection{Morphological evolution of dust and ionized gas emission}
In Fig.~\ref{fig:RGB_dust_free} we show four-panels emission maps corresponding to the  {\sc RAMSES} snapshots S$_3$-S$_6$ illustrating the temporal evolution of dust emission and ionized gas around the massive ionizing source. Each map combines $70\ \mu$m emission (green), tracing thermal radiation from hot dust associated with recent star formation, $250\ \mu$m emission (red) from colder dust, and $1.3\ $GHz emission (blue), tracing free-free radiation from ionized gas.

Snapshot S$_3$ represents an early compact evolutionary stage. The stellar cluster appears in yellow as it is surrounded by bot hot and cold dust  is spatially offset from the ionizing star by a projected distance of order a few cluster radii. A compact ($\lesssim 2 - 3\ \mathrm{pc}$) region of enhanced $70\ \mu$m emission surrounds the ionizing source, indicating localized dust heating. Nevertheless, the lack of spatially extended $1.3\ \mathrm{GHz}$ emission on that map implies that the ionized volume fraction within the central cavity is still small, and no developed H{\sc\,ii} region is present on parsec scales.

During S$_4$, the large-scale morphology of the $250\ \mu\mathrm{m}$ dust emission remains nearly unchanged, suggesting that the bulk mass distribution of cold dust and dense gas is still largely intact. The hot-dust component expands moderately, increasing its characteristic radius by a factor of $\sim 1.5$ relative to S$_3$. Despite this thermal response, the surrounding gas structures show only minor displacement, indicating that radiation feedback has not yet driven strong dynamical expansion. Free-free emission remains confined to the immediate vicinity of the ionizing star.

A clear dynamical transition occurs around the time of S$_5$ where the H{\sc\,ii} regions enters the extended phase. Gas ionization become prominent, and the central density cavity grows as material is pushed outward. At the same time, the $1.3\ \mathrm{GHz}$ emission increases substantially and fills a region comparable in size to the cavity, implying that the ionized region now occupies a significant fraction of the central volume. This stage marks the formation of a resolved H{\sc\,ii} region whose radius parsec scales where two distinct lobes start to appear.

By snapshot S$_6$, ionization feedback dominates the central morphology. The H{\sc\,ii} region has expanded into the stellar-cluster environment, and the cavity appears almost entirely ionized. Strong, spatially extended free-free emission fills most of the evacuated volume, while the thermal dust emission primarily traces dense shells and clumps at the periphery. The evolution from S$_3$ to S$_6$ therefore reflects a transition from a dust-heating-dominated phase to one in which ionization-driven expansion governs the gas dynamics on parsec scales. We note that, in this evolutionary phase, the H{\sc\,ii} region is no longer symmetric but has developed two lobes and a central constriction (see also Fig.~\ref{fig:column_density}). Furthermore, at this stage, outflows and pillar-like structures, formed as a result of stellar feedback, are driven by stars embedded in the shell structure.

\subsection{Evolution of AME and free-free emission}
In Fig.~\ref{fig:RGB_AME_free} we show the same set of {\sc RAMSES} snapshots (S$_3$ - S$_6$) as in Fig.~\ref{fig:RGB_dust_free}, but now focusing on microwave tracers. The maps show AME at the peak frequency (orange) overlayed free-free emission at $1.3$\,GHz (blue). In addition to the ionizing star, the complete stellar component of the cluster is indicated by white dots.

The two snapshots S$_3$ to S$_4$  correspond to early compact stages in which the emission morphology is already governed by the small but expanding H{\sc\,ii} region. Both free-free and AME are centrally concentrated, with characteristic radii well below a parsec. The AME peak frequency (also shown as orange line in Fig.~\ref{fig:SEDs_aperture}) lies in the range $\nu_{\mathrm peak} \approx 11 - 26\ \mathrm{GHz}$, consistent with rapidly rotating NGs exposed to a strong local radiation field.

By S$_5$, the free-free emission becomes detectable on parsec scales as the H{\sc\,ii} region expands and the central cavity grows. The $1.3$\,GHz morphology now clearly traces the ionized bubble interior. AME also follows the bubble, but covers a slightly larger projected area than the free-free emission, extending into the transition layer between ionized and neutral gas. This indicates that spinning dust grains remain abundant in the partially processed shell where gas densities and radiation fields jointly favor efficient rotational excitation. The spatial coincidence of AME enhancements with the stellar cluster in the background suggests that the cluster radiation field raises the rotational temperature of the smallest grains to values comparable to those near the ionizing star. As a consequence, AME structures are visible not only around the ionizing source but also in the vicinity of the cluster, including components peaking near $\sim 25$\,GHz.

Only minor spectral evolution occurs between the extended phases S$_5$ and S$_6$, with the peak AME frequency shifting slightly to $\nu_{\mathrm peak} \simeq 26.0$\,GHz. At this stage, the H{\sc\,ii} region interior is here prominently visible in free-free emission, whereas the AME predominantly arises from the dense shell surrounding the ionized cavity. The spatial segregation between blue and orange components therefore highlights a transition from ionized gas in the bubble interior to spinning-dust emission in the compressed, photo-processed shell.

\subsection{LOS effects on the free-free SED}
In Fig.~\ref{fig:SED_free} we display the spectral energy distribution (SED) of the free-free emission  in snapshot S$_1$, corresponding to the earliest stage of its evolution equivalent to an ultra compact H{\sc\,ii} region stage \citep{Habing1979,Wood1989}. The spectra are derived for three orthogonal lines of sight (LOS) along the $x$-, $y$-, and $z$-directions of the {\sc RAMSES} AMR grid and are integrated within an aperture defined by the spatial distribution of the electron density as shown in Fig.~\ref{fig:column_density} and listed in Table~\ref{tab:MHD_param}.

Classical thermal bremsstrahlung theory predicts a well-defined transition between two spectral regimes. At low frequencies, where the plasma becomes optically thick ($\tau_\nu \gtrsim 1$), the flux density scales as $S_\nu \propto \nu^{2}$. At higher frequencies, once the emission becomes optically thin ($\tau_\nu \ll 1$), the spectrum approaches $S_\nu \propto \nu^{-0.1}$. These slopes, however, strictly describe an idealized, homogeneous medium with constant physical conditions along a single LOS.

In contrast to such ideal conditions, the simulated SEDs depart from these canonical power-laws because each LOS intersects multiple density structures along its path characterized by different emission measures and temperatures. Moreover SEDs are the collected flux within the characteristic radius of the ionized gas. However, even for a narrow beam, the observed flux would represents the superposition of emission from regions spanning a range of optical depths. This LOS mixing broadens the turnover between optically thick and thin regimes and produces effective spectral slopes that vary with viewing direction.

Differences between the $x$-, $y$-, and $z$-spectra are modest in the optically thin regime, where emission adds approximately linearly and becomes less sensitive to local density contrasts. Toward lower frequencies, optical depth effects become dominant and the divergence between the LOS increases, reflecting anisotropies and clumpiness in the density structure of the young H{\sc\,ii} region. Such behavior demonstrates that unresolved inhomogeneities can significantly modify the free-free SED compared to the textbook homogeneous case. This effect must be considered when performing SED fitting, particularly for observations with limited spectral coverage, where assuming a single turnover frequency or a uniform spectral slope may lead to biased physical parameters.

\subsection{Multi-Component SED during the H{\sc\,ii} region expansion}
\label{sect:SED}
Figure~\ref{fig:SEDs_aperture} presents six panels showing the SEDs of the post-processed {\sc RAMSES} snapshots S$_1$ - S$_6$ over the frequency range $\nu = 1 - 10^{4}\ \mathrm{GHz}$. Each panel decomposes the total emission into free-free radiation, thermal dust emission from {\sc Astrodust}, and AME. For comparison, the data points representing the SEDs derived from Planck, WMAP, and ground-based facilities  of a source in Perseus and of a pure H{\sc\,ii} region in Orion are overplotted\footnote{We note that uncertainties for the observed SEDs are not provided in the corresponding publications.} in each panel \citep{Planck2011,CollaborationPlanck2012}. 

Because the distance of the post-processed snapshot region to the observer is a free parameter in the RT modeling, the observational SEDs are rescaled in amplitude to match the overall flux level of our synthetic spectra. All model fluxes are integrated within an aperture defined by the electron-density radius of the respective snapshot (see Table~\ref{tab:MHD_param}).

Again, the S$_1$ panel corresponds to the youngest evolutionary and ultra-compact phase. At this time, the aperture encompasses primarily the immediate surroundings of the ionizing star. Dust temperatures are therefore elevated compared to the reference observations, and the thermal dust SED peaks in the mid-infrared. The free-free component exhibits both optically thick and optically thin contributions, displaying the characteristic turnover from $S_\nu \propto \nu^{2}$ at low frequencies to $S_\nu \propto \nu^{-0.1}$ at higher frequencies (see Fig.~\ref{fig:SED_free}). The AME component peaks at $\nu \approx 37$\,GHz, comparable to the H{\sc\,ii} region modeled in \cite{AME_Planck2016}. We emphasize again that at this stage, the spatial origin of the peak AME largely spacialy coincides with the compact free-free emitting region.

As the H{\sc\,ii} region expands from S$_2$ to S$_5$, the aperture encloses progressively larger volumes containing both hot and colder gas and dust components (compare Fig.~\ref{fig:RGB_dust_free}). The increasing contribution of cooler dust within the aperture shifts the thermal peak toward the far-infrared, bringing the synthetic SED into closer agreement with the Planck reference spectra. Concurrently, the free-free emission becomes increasingly optically thin as the density in the ionized region decreases. The spectral slope gradually approaches $S_\nu \propto \nu^{-0.1}$ across a broad frequency interval.

By S$_6$, the free-free emission is optically thin over the entire considered frequency range, resulting in an approximately constant slope consistent with $S_\nu \propto \nu^{-0.1}$. The AME amplitude initially declines from S$_1$ to S$_3$ during the intermediate stages. This is not a result of the larger aperture but because the ionized interior becomes more tenuous, reducing the efficiency of rotational excitation within the central cavity. However, the continued growth of the aperture introduces additional contributions from the stellar cluster in behind the H{\sc\,ii} as well as surrounding neutral material. Consequently, the AME intensity increases again in S$_6$, and the peak frequency shifts toward $\nu_{\mathrm peak} \simeq 26.0$\,GHz. 

Overall, the sequence S$_1$-S$_6$ illustrates how the interplay between ionization-driven expansion, evolving dust temperatures, and changes in grain excitation conditions shapes the composite radio-to-infrared SED of a developing H{\sc\,ii} region.

In Fig.~\ref{fig:SEDs_no_aperture} the same set of SEDs as presented in Fig.~\ref{fig:SEDs_aperture} for the snapshots S$_1$ - S$_6$, but now integrated over the entire $8\ \mathrm{pc} \times 8\ \mathrm{pc}$ emission map rather than being restricted to the aperture defined by the electron-density radius. This global integration allows us to assess how extended neutral material modifies the composite spectrum.

Because the free-free emission originates exclusively from the ionized gas in the expanding H{\sc\,ii} region, its magnitude and spectral turnover remain essentially unchanged compared to the aperture-restricted case shown in Fig.~\ref{fig:SEDs_aperture}. Naturally, the characteristic transition from the optically thick regime ($S_\nu \propto \nu^{2}$) to the optically thin regime ($S_\nu \propto \nu^{-0.1}$) occurs at the same frequencies, reflecting the fact that no additional ionized material is included outside the original aperture.

In contrast, the thermal dust emission is broader with a plateau at its peak once the full region is considered. Contributions from both hot dust near the ionizing source and colder dust in the surrounding molecular environment combine to produce a wider SED. The effect is particularly pronounced for S$_1$, where the ionizing star is still embedded in dense material. Strong local heating of nearby dust grains results in an extended high-frequency tail and a comparatively broad infrared peak. By S$_6$, the ionizing star resides inside a well-developed density cavity. The harder radiation field is distributed over a larger volume, leading to a more moderate temperature contrast and resulting in a correspondingly narrower dust SED.

The AME component exhibits a more complex behavior. In addition to the main peak—associated with spinning grains in the evolving H{\sc\,ii} region and in the stellar-cluster environment, a secondary peak emerges at lower frequencies. This additional component originates from NGs in the more diffuse material outside the ionized cavity, where lower gas densities and weaker radiation fields shift the rotational emissivity toward smaller characteristic frequencies. The presence of two distinct AME maxima therefore reflects the coexistence of dense, strongly irradiated gas and more extended diffuse material within the entire simulation domain.

Our synthetic SEDs of the different evolutionary stages of H{\sc\,ii} 
(S$_1$-S$_6$) agrees well with what is actually observed in multple Galactic and extragalactic sources \citep[e.g.][]{Planck2014}. In molecular and dark clouds, the gas is predominantly neutral, with moderate radiation fields and temperatures of $T_{\mathrm{g}}\sim 10\ \mathrm{K}-40\ \mathrm{K}$ \citep{Traficante15, Elia2017, Elia2021}. In contrast, classical H{\sc\,ii} regions exhibit a combination of physical conditions that efficiently suppress spinning dust emission and force rotational disruption of grains. Hence, the reduction in nano-grains leads to reduced emissivity at microwave frequencies \citep{DraineLazarian1998,Hoang2016}. Furthermore, the continuum emission from ionized gas in H{\sc\,ii} regions is dominated by free-free radiation, which rises toward lower frequencies and can completely mask any residual spinning dust component. Spectral analyses of prototypical regions such as the Orion Nebula (M\,42) and NGC\,6357 show that their microwave spectra are well reproduced by pure free-free plus thermal dust emission without requiring an additional AME component  \citep{AME_HIIfree2010,Gonzales2025}. Together, these effects explain why AME is strong in molecular and dark clouds, where NGs are abundant and survive in neutral gas, but is weak or absent in ionized environments where those carriers are destroyed or their emission overwhelmed by free-free radiation.

\subsection{LOS variations in emission and optical depth}
Since the H{\sc\,ii} region is not spherically symmetric but, particularly in the case of S$_6$, exhibits two lobes and a constriction, in the center we examine in this section the effects of different LOSs.

In Fig.~\ref{fig:SED_LOS} we present four panels showing the SEDs of snapshots S$_1$ and S$_6$, including free-free, AME, and thermal dust emission, together with their corresponding optical depths. As in Fig.~\ref{fig:SED_free}, the spectra are displayed for the emission along the three different LOSs corresponding to the principal axis of the {\sc RAMSES} snapshots. Dashed lines indicate the mean flux, while the shaded regions represent the full range of synthetic spectral fluxes obtained from the three different orthogonal viewing angles.

For snapshot S$_1$, the free-free emission exhibits nearly identical fluxes at $\nu \sim 10^{4}$\,GHz, where the plasma is optically thin along all LOS. Toward lower frequencies, the spread between LOS increases markedly. Around $\nu \sim 10^{2}$\,GHz the transition toward optical thickness begins, and at $\nu \sim 1$\,GHz the optical depth can reach values as high as $\tau_\nu \approx 10^{4}$ along particular LOSs. In this regime, parts of the ionized region become effectively shielded, and the observed flux becomes strongly dependent on the density structure intersected by a given LOS. 

A different behavior is found for S$_6$. Here, the H{\sc\,ii} region is optically thin across the entire frequency range considered. Consequently, both the mean free-free spectrum and the LOS spread remain narrow, and the flux shows only minor directional variations.

Thermal dust emission displays the opposite trend with respect to frequency. The dispersion between LOS increases toward higher frequencies, where dust absorption becomes significant. Optical depths reach values up to $\tau_\nu \approx 10$ in the far-infrared for S$_1$, producing a broader range of observed fluxes at high $\nu$. Unlike the free-free case, optically thick regimes for dust are present throughout the evolutionary sequence from S$_1$ to S$_6$. As a result, LOS effects remain important for dust observations during the entire expansion of the H{\sc\,ii} region.

AME does not possess an intrinsic absorption coefficient (see Sect.~\ref{sect:ame_emission}). However, it can be attenuated by free-free or dust absorption along the LOS. Dust absorption becomes relevant only at $\nu \gtrsim 8\times10^{3}\ \mathrm{GHz}$, well above the AME peak frequencies, so the two spectra components do not overlap (see also Sect.~\ref{sect:SED}  for details). Free-free absorption may affect the low-frequency tail of the AME during the earliest stage, since in S$_1$ the ionized gas is optically thick for $\nu \lesssim 10$\,GHz. Although this would, in principle, suppress the low-frequency AME component, modeling such coupled RT effects lies beyond the scope of the present approach. In practice, the free-free emission in this regime exceeds the AME by approximately $0.5-1\ \mathrm{dex}$, implying that any absorption-modified AME contribution at low frequencies would remain observationally negligible.

\subsection{Impact of rotational disruption on thermal dust emission and anomalous  microwave emission}
In Fig.~\ref{fig:dust_RD}, we compare the SEDs of thermal dust emission, free-free emission, and AME for snapshots S$_1$ and S$_6$. The SEDs are computed for dust grain with three distinct values of the maximum tensile strength, $\mathcal{S}_{\mathrm{max}}=1.0\times10^{8}\ \mathrm{N\,m}^{-2}$, $1.0\times10^{7}\ \mathrm{N\,m}^{-2}$, and $1.0\times10^{6}\ \mathrm{N\,m}^{-2}$, respectively.

In snapshot S$_1$, the ionizing star is still embedded in its dense natal cloud, where a substantial fraction of the UV radiation is absorbed and reprocessed into the IR regime. Consequently, RATs are less efficient at spinning up dust grains, and RD is nearly absent in this snapshot (compare Fig.~\ref{fig:tdust}). As a result, the SEDs are essentially independent of the adopted value of $\mathcal{S}_{\mathrm{max}}$.

In contrast, in snapshot S$_6$, the ionizing star is surrounded by an ionized cavity containing low-density gas, allowing UV radiation to propagate farther through the simulation domain. LGs spun up to high angular velocities by RATs can undergo efficient RD, particularly in the vicinity of the ionizing star, where the radiation field is strongest. As a result, the thermal dust SED is modestly modified. The inset in the right panel of Fig.~\ref{fig:dust_RD} illustrates a slight decrease in thermal emission for lower values of $\mathcal{S}_{\mathrm{max}}$, as a fraction of the large grains contributing to the IR emission is disrupted. Although the overall change in thermal emission is small, it is systematic across the grain size distributions considered here.

In our grain-redistribution model, the transfer of mass from the large-grain end of the size distribution towards NGs has two consequences. First, it increases the total number of NGs. Second, it enriches the  H{\sc\,ii} region with very small NGs. The increased abundance of NGs enhances the AME, with the model adopting $\mathcal{S}_{\mathrm{max}}=1.0\times10^{6}\ \mathrm{N\,m}^{-2}$ producing more AME flux at slightly higher frequencies than the model adopting $\mathcal{S}_{\mathrm{max}}=1.0\times10^{8}\ \mathrm{N\,m}^{-2}$.

Moreover, the angular velocity distribution of NGs is approximated by a Maxwellian distribution (see Sect.~\ref{sect:ame_emission}). Hence, RD preferentially removes grains from the high-angular-velocity tail of this distribution. Consequently, the high-frequency tail of the AME spectrum, associated with the fastest rotating grains, is suppressed at about $\nu = 100\ \mathrm{GHz}$. In the models shown here, the AME amplitude varies with $\mathcal{S}_{\mathrm{max}}$, whereas the position of the main emission peak is only weakly affected. The high-frequency cutoff of the AME spectrum may therefore provide a means of constraining $\mathcal{S}_{\mathrm{max}}$ through multi-frequency observations.

Furthermore, we report for the other destruction mechanisms considered in our RT modeling are comparatively inefficient in the presented evolutionary sequence S$_1$ - S$_6$. Thermal sputtering is usually considered a major factor in the grain size distribution ans it becomes most effective at high gas temperatures. However, the time intervals between successive {\sc RAMSES} snapshots are too short to substantially redistribute grain sizes and to produce a measurable change in the SED. Dust sublimation and Coulomb explosions also occur in our post processing step but affect only a very small fraction of cells ($\sim 1\times 10^{-3} - 2\times 10^{-3}\ \%$ per snapshot), and therefore do not significantly alter the global thermal emission and the AME.

Taken together, these results indicate that RD is the dominant dust-destruction mechanism shaping the emergent SED in our synthetic images. It modifies both the infrared thermal emission through the depletion of large grains and the AME component through the suppression of rapidly rotating NGs while other processes remain subdominant over the considered evolutionary timescales.
\begin{figure*}[]
	\centering
	\includegraphics[width=1.0\textwidth]{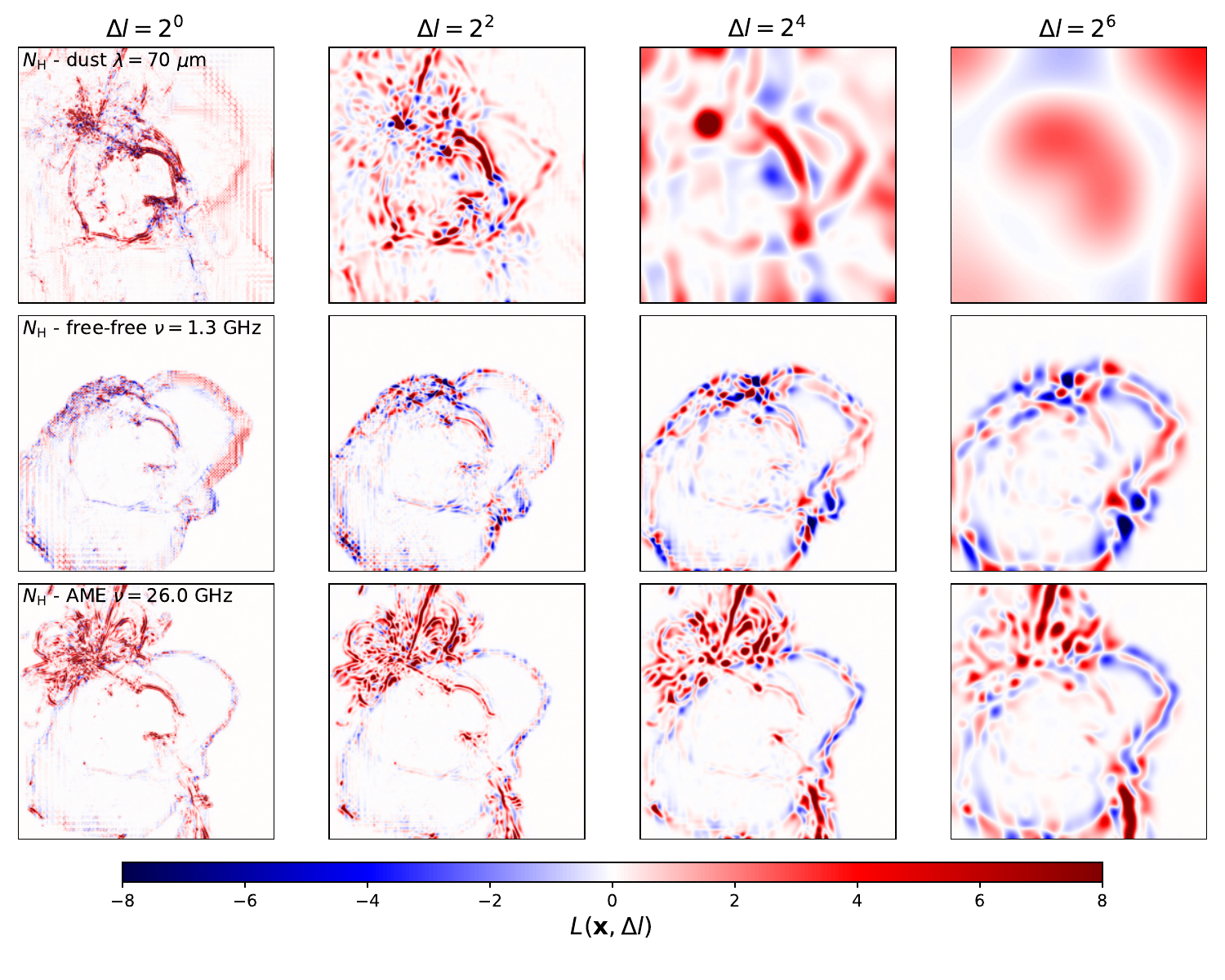}
	\caption{ The local correlation function $L(\mathbf{x},\Delta l)$ (see appendix~\ref{app:multiscale}) quantifying the alignment of structures in the S$_6$ maps of gas column density $N_{\mathrm{H}}$ and the ones of $70\ \mu\mathrm{m}$ dust emission (top row), $1.2\ \mathrm{GHz}$ free-free emission (middle row), and the $26.0\ \mathrm{GHz}$ peak AME  on scales with a dimensionless length scale $\Delta l=2^0 - 2^6$ (left to right columns). }
	\label{fig:corr_NH}
\end{figure*}

\begin{figure*}[]
	\centering
	\includegraphics[width=1.0\textwidth]{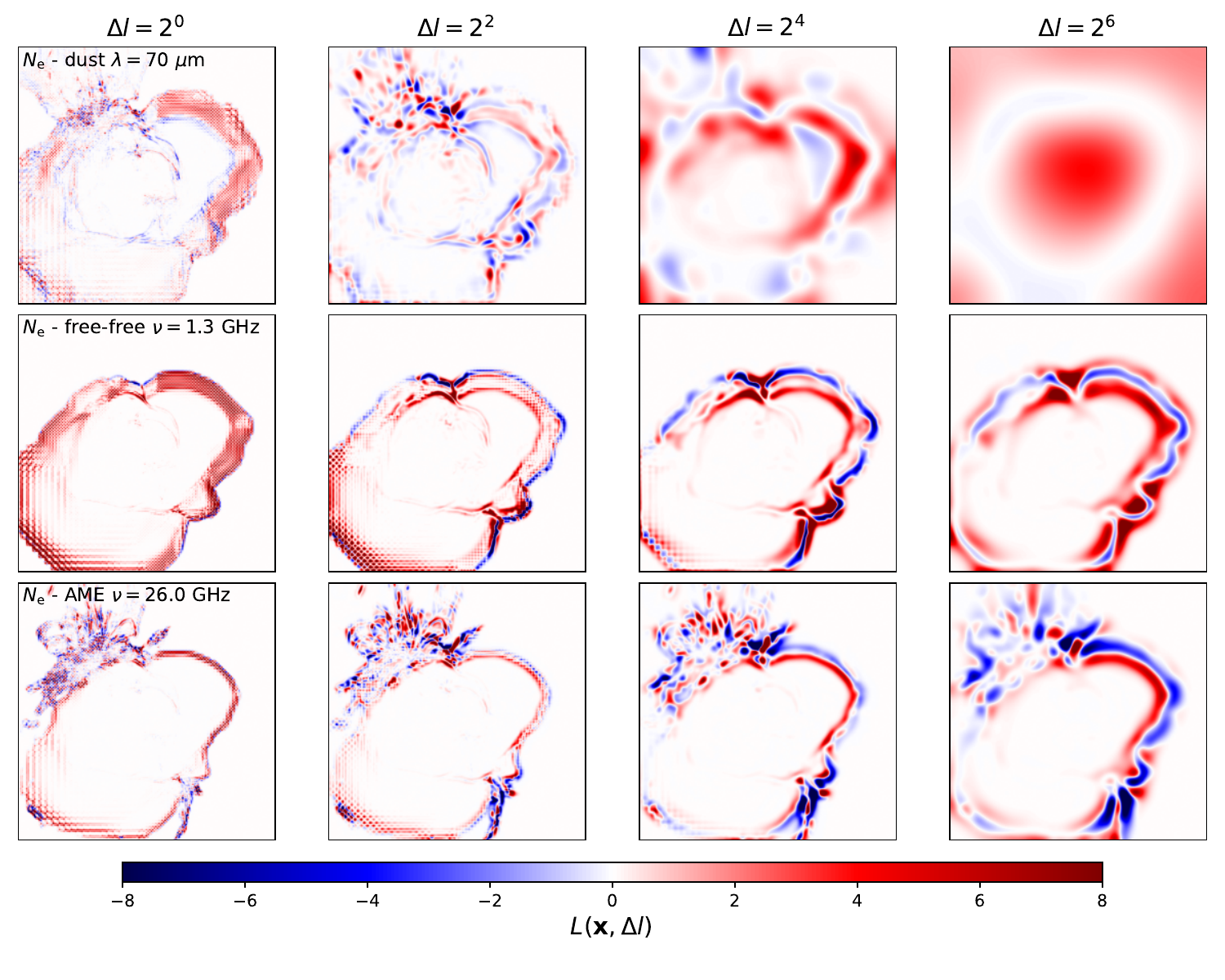}
	\caption{ The same as Fig.~\ref{fig:corr_NH} but for the alignment of structures seen in the electron column density $N_{\mathrm{el}}$ and the emission maps.}
	\label{fig:corr_Nel}
\end{figure*}

\begin{figure}[]
	\centering
	\includegraphics[width=0.49\textwidth]{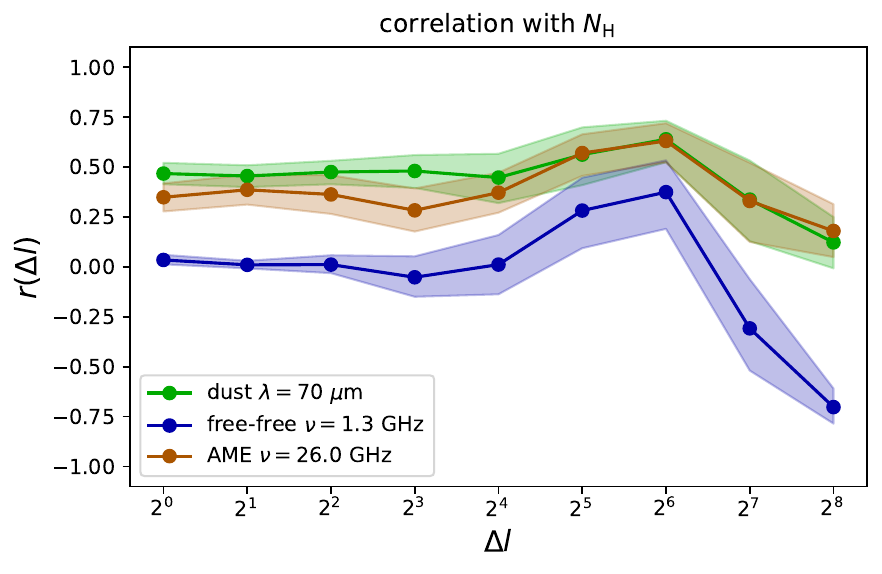}\\
    \includegraphics[width=0.49\textwidth]{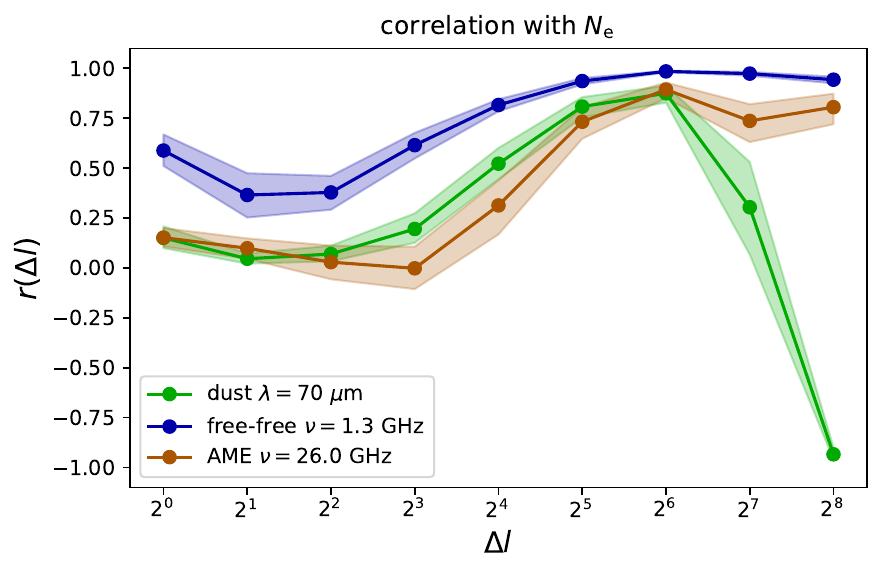}
	\caption{ The global correlation coefficient $r(\Delta l)$ on characteristic scales $\Delta l = 2^0 - 2^8$. Considered are the correlations of the $70\ \mu\mathrm{m}$ dust emission (green), $1.2\ \mathrm{GHz}$ free-free emission (blue), and the $26.0\ \mathrm{GHz}$ peak AME (orange) with the gas column density $N_{\mathrm{H}}$ (top panel) and the electron column density $N_{\mathrm{el}}$ (bottom panel), respectively. }
	\label{fig:corr_r}
\end{figure}
\subsection{Multi-scale correlation of emission and column density}
Since both free-free and dust-related emission depend on multiple physical parameters 
(e.g.\ density, temperature, radiation field strength, and grain properties) this raises the questions what physical quantities  are traced exactly by spatially resolved emission observations of  H{\sc\,ii} regions. The aim of this section is to quantify how the spatially resolved flux correlates with the total gas column density $N_{\mathrm{H}}$ and the electron column density $N_{\mathrm{el}}$ as a function of spatial scale $\Delta l$ by performing a multi-scale structure correlation analysis (see appendix~\ref{app:multiscale} for methodological details). 
  
In Fig.~\ref{fig:corr_NH} we present maps of the local correlation function $L(\vec{x},\Delta l)$ for snapshot S$_6$ between the emission components and $N_{\mathrm{H}}$. On the smallest resolved scale ($\Delta l = 1$), the $70\ \mu\mathrm{m}$ dust emission exhibits a slightly positive correlation with the edges of the central cavity of the H{\sc\,ii} region, indicating that hot dust traces merely the compressed and irradiated gas at the ionization front. No significant alignment is found 
with the full shell structure at this scale. As $\Delta l$ increases, a positive signal 
emerges in the shell and in the dense structures surrounding the background star cluster, reflecting the increasing contribution of  dust associated with higher column densities.

The free-free emission shows a distinct behavior. The correlation signal traces both the central cavity as well as the shell of the H{\sc\,ii} region. However, positive and negative contributions to $L(x,\Delta l)$ appear in roughly equal measure across all scales. This mixed signal indicates that free-free emission does not simply scale with total gas column density but instead depends sensitively on the ionization structure. No statistically significant alignment of structures is found with the star-cluster.

In contrast, the AME exhibits a predominantly positive alignment with $N_{\mathrm{H}}$ in the central cavity, the outer shell, and the star-cluster region across nearly all spatial scales. Toward larger $\Delta l$, the fraction of anti-aligned structures increases, suggesting that diffuse material begins to dominate the large-scale signal and partially decouples AME from the densest gas components.

In Fig.~\ref{fig:corr_Nel} we show the corresponding correlation maps between the emission components and the electron column density $N_{\mathrm{el}}$. The thermal emission from hot dust shows a positive alignment with both the central cavity and the outer shell, consistent with enhanced heating near ionized gas. At the same time, significant anti-aligned regions appear on multiple scales, reflecting locations where dust emission originates from neutral material with little associated free electrons.

The free-free emission aligns with the shell on small scales, as expected from the 
ionized boundary layers. With increasing $\Delta l$, anti-alignment becomes prominent 
along the ridges of the bipolar lobes, indicating that geometrical dilution and density 
gradients dominate the large-scale signal.

Although AME could, in principle, correlate with $N_{\mathrm{el}}$ through the dependence of grain charging on electron density, the simulations show that this effect is minor compared to the intrinsic dipole moment of the NGs. A weak positive correlation is detected on small scales coinciding with the lobes, but no correlation is found with the central cavity. Instead, an anti-correlation appears in the star-cluster region. At larger $\Delta l$, the lobe structures also transition toward anti-correlation, further demonstrating that AME is primarily governed by grain properties and local radiation fields rather than by the electron column density itself.

Fig.~\ref{fig:corr_r} summarizes the cross-correlation parameter $r(\Delta l)$ (see appendix~\ref{app:multiscale}) as a function of characteristic scale over the range $\Delta l = 2^{0}$-$2^{8}$, quantifying the average alignment of structures between the synthetic emission maps and the gas column density $N_{\mathrm{H}}$ as well as the electron column density $N_{\mathrm{el}}$, respectively. With respect to $N_{\mathrm{H}}$, the hot-dust tracer at $70\ \mu\mathrm{m}$ shows a consistently positive correlation across all scales, with $r(\Delta l)\simeq 0.5$ and only weak scale dependence. The AME correlation behaves similarly but with a lower overall amplitude, $r(\Delta l)\simeq 0.25$, and a more pronounced decline toward the largest scales ($\Delta l=2^{8}$), indicating that the large-scale AME morphology becomes progressively less coupled to the total gas column. In contrast, free-free emission exhibits no significant correlation with $N_{\mathrm{H}}$ on small to intermediate scales ($r(\Delta l)\simeq 0$ up to $\Delta l=2^{4}$), and transitions to an anti-correlation on larger scales, reaching $r(\Delta l)\simeq -0.75$ at $\Delta l=2^{8}$.

For correlations with the electron column density $N_{\mathrm{el}}$, the expected strong coupling of free-free emission to ionized gas is recovered. The correlation is already high on the smallest scales, $r(\Delta l)\simeq 0.5$, followed by a local minimum around $\Delta l=2^{1}$, and then rises again toward large scales, approaching $r(\Delta l)\rightarrow 1$ for $\Delta l\gtrsim 2^{4}$-$2^{8}$. AME is less strongly correlated with $N_{\mathrm{el}}$ at small scales ($r(\Delta l)\simeq 0.25$), but the correlation increases steadily with scale and reaches $r(\Delta l)\simeq 0.75$ at the largest $\Delta l$, consistent with an increasing contribution from extended partially ionized structures. The hot-dust emission shows a comparable level of correlation to AME on small and intermediate scales, but turns into a strong anti-correlation on the largest scales, suggesting that the large-scale dust-heating morphology becomes decoupled from the distribution of free electrons once the analysis averages over both ionized cavities and extended neutral material. Interestingly, the decoupling between hot dust emission and electron density occurs at a scale of $\Delta l \approx 2^{6}$, corresponding to $\approx 1$ pc. Remarkably, this is the same scale at which the gas kinematics have also been found to decouple, transitioning from a likely turbulence-dominated regime on scales above $\sim 1$ pc to a gravity (and feedback) dominated regime below $1$ pc \citep{Traficante2018, Traficante2020, Benedettini21, Peretto2023, Colman24}. The decline in the correlation between hot dust and $N_{\mathrm{el}}$ on scales larger than $1$ pc may therefore indicate that the impact of H,\textsc{ii}  region feedback becomes decoupled from the global collapse of the region. Above this scale, turbulence may instead dominate the gas kinematics, largely independently of both the strength of the free-free emission and the gravitational content of the region.

The correlation patterns reported above can be understood in the context of the distinct physical dependencies of free-free, AME, and thermal dust emission. Free-free emission scales with the emission measure, $\mathrm{EM}=\int n_{\mathrm{el}}^2\ \mathrm{d}\ell$, and therefore depends quadratically on the electron density rather than directly on the total gas column density $N_{\mathrm{H}}$. As a consequence, strong positive correlations with $N_{\mathrm{el}}$ are expected in ionized boundary layers and compressed shells where $n_{\mathrm el}$ is enhanced, while mixed or even anti-correlated signals with $N_{\mathrm{H}}$ naturally arise in regions where large neutral columns coexist with little ionized gas. Thermal dust emission, in contrast, is proportional to the dust density weighted by the Planck function (see Sect.~\ref{sect:dust_emission}). Its correlation with $N_{\mathrm{H}}$ therefore reflects both the amount of material and the local dust temperature. Positive correlations with the shell and dense structures are expected where heating and column density coincide, whereas anti-correlations emerge in hot but tenuous cavities or in cold, shielded regions with weak radiation fields. The behavior of AME can be attributed to the microphysics of rapidly rotating nano-grains. The emissivity depends on the abundance of NGs and on their rotational excitation, which is dominated by a combination of gas collisions and IR emission. Although grain charging depends on $n_{\mathrm el}$, the resulting modification of the electric dipole moment is typically subdominant compared to the intrinsic dipole moment of the {\sc Astrodust} NGs. Consequently, AME correlates more strongly with total $N_{\mathrm{H}}$ tracing the spatial distribution of NGs than with $N_{\mathrm{el}}$. The emergence of anti-correlations on larger scales likely reflects geometrical dilution and the coexistence of diffuse neutral material with low excitation efficiency of NGs. Overall, the multi-scale correlation structure directly mirrors the different density scalings, temperature sensitivities, and excitation mechanisms underlying each emission process.

We emphasize that, within our radiative transfer modelling, the thermal dust emission is computed independently of the electron distribution. The dust temperature and resulting infrared emission are determined by the local radiation field and dust properties, and do not explicitly depend on $N_{\mathrm{el}}$. Nevertheless, ionization and the formation of the H{\sc\,ii} region shell are both governed by feedback from the ionizing star. This feedback simultaneously redistributes the gas and sets the electron density structure. Consequently, gas and electrons and hence dust emission are not completely physically independent components of the system.

As a result, correlations between thermal dust emission and $N_{\mathrm{el}}$ arise indirectly. The expanding ionization front compresses material into a dense shell, where both the electron density, at least locally in partially ionized layers, and the dust temperature can be enhanced due to strong irradiation. In addition, geometrical effects couple the morphology of hot dust emission to the structure of the ionized cavity and its surrounding shell. The measured correlation between dust emission and electron column density therefore reflects their common origin in stellar feedback and large-scale gas dynamics, rather than a direct microphysical dependence of dust emissivity on the electron distribution.

\section{Outlook}
The present work provides idealized synthetic observations at full spatial and spectral resolution. A natural next step is to include instrumental effects in order to generate realistic mock observations tailored to specific facilities \citep{Nucara25}. Convolving the synthetic maps with interferometric response functions, uv-coverage, beam dilution, missing short spacings, spectral sampling, calibration uncertainties, and realistic noise properties would allow direct comparison with observations from instruments such as ALMA (probing the FIR/sub-mm dust component and compact ionized structures) and the upcoming SKA (probing free-free and AME emission at centimeter wavelengths).

While ALMA mock observations have been produced and analyzed in \citep{Nucara25} to explore clump fragmentation properties as determined from the 1.3\,mm continuum emission, the first SKA mock observations have been presented in \citet{Traficante26}, but will be analyzed in a upcoming work that will help determine the conditions under which H{\sc\,ii} regions can be detected, particularly when they are faint and in their earliest evolutionary stages. By combining the spatial extent of H{\sc\,ii} regions recovered from SKA and the continuum fragmentation properties recovered from ALMA mock observations with the simulated population of sink particles, we will be able to investigate whether the expansion of ionized regions correlates with triggered star formation in the surrounding gas or influences the properties of ongoing star formation.

These mock observations will also enable us to assess under which conditions AME can be disentangled from free-free emission in evolving H{\sc\,ii} regions.

Dust polarization is not considered in the present study. However, polarization is expected for both thermal dust emission and AME \cite{LazarianHoang2007,Hoang2016ApJ}. Hence, polarization observations could provide complementary constraints on the magnetic-field geometry and grain-alignment properties within H{\sc\,ii} regions. Such measurements represent a promising avenue for future observational programmes with ALMA and the SKA, which can, in principle, probe polarized dust emission
at millimetre wavelengths and polarized AME at centimetre wavelengths, respectively, thereby maximizing the diagnostic potential of multi-frequency continuum studies of star-forming regions.

\section{Summary}
\label{sect:summary}
In this work, we combined three-dimensional radiation-magnetohydrodynamic (RMHD) simulations with detailed radiative transfer (RT) post-processing to predict the multi-frequency continuum emission of a star-forming cloud evolving into an H{\sc\,ii} region. The dynamical evolution was computed with the adaptive mesh refinement code {\sc RAMSES}, including self-gravity, turbulence, magnetic fields, sink particle formation, protostellar jets, and ionizing radiation feedback. From this simulation sequence, we selected six representative snapshots (S$_1$--S$_6$) tracing the transition from a compact, embedded ionizing source to an extended, asymmetric H{\sc\,ii} region with bipolar lobes and shell structures.

The snapshots were post-processed with the Monte Carlo RT code {\sc POLARIS}, extended to self-consistently treat (i) thermal emission from composite {\sc Astrodust} grains, (ii) free-free emission from ionized gas, and (iii) anomalous microwave emission (AME) from rapidly rotating nano-grains (NGs). The dust size distribution consists of an MRN-type distribution for large grains (LGs) and a log-normal distribution for NGs. The effective material properties of {\sc Astrodust}, including its intrinsic electric dipole moment, account for the constituent materials of the composite grains. Grain charging is computed using a steady-state equation that includes collisional charging, photoelectric emission, and photodetachment. We additionally account for dust destruction through rotational disruption (RD) driven by radiative torques (RATs), Coulomb explosion, sublimation, and thermal sputtering, allowing us to assess feedback-induced modifications of the grain size distribution and the resulting emission.

Synthetic emission maps and spectral energy distributions (SEDs) were generated over a broad frequency range (${ 1-10^{4}\ \mathrm{GHz} }$), considering different lines of sight (LOS), apertures, and full-domain integrations. We further quantified the structural coupling between the emission components and the total gas column density $N_{\mathrm{H}}$ and electron column density $N_{\mathrm{el}}$ using a multi-scale correlation analysis.

We summarize our major findings as follows:

\begin{itemize}

\item The evolution from an ultracompact region (S$_1$), with a spatial extent of $0.06\ \mathrm{pc}$, to an extended region (S$_6$), with a spatial extent of approximately $4.9\ \mathrm{pc}$, reveals a transition from a compact, dust-heating-dominated configuration to one in which ionizing feedback strongly shapes the emission morphology. Thermal dust emission initially traces heated material surrounding the embedded ionizing source, whereas free-free emission at later stages delineates an expanding, asymmetric H{\sc\,ii} region with lobes and shell structures.

\item Early snapshots exhibit a characteristic turnover from optically thick ($S_\nu \propto \nu^{2}$) to optically thin ($S_\nu \propto \nu^{-0.1}$) free-free emission. As the H{\sc\,ii} region expands and its density decreases, the free-free spectrum becomes optically thin over the considered frequency range. LOS-dependent variations are significant during the optically thick phase but become negligible once the region is optically thin.

\item The AME peak frequency evolves from approximately $20$--$25\ \mathrm{GHz}$ in the earliest stage to approximately $26\ \mathrm{GHz}$ in the evolved configuration. Initially, the peak AME is spatially coincident with the compact ionized region. At later stages, AME predominantly traces the compressed shell and irradiated cluster environment. Secondary low-frequency AME components arise from diffuse material when the emission is integrated over the full simulation domain.

\item Thermal dust SEDs depend strongly on the chosen aperture. When the integration is restricted to the electron-density radius, the thermal-emission peak shifts from the mid-IR towards the far-IR as the H{\sc\,ii} region expands. Full-domain integration broadens the SED through the combined contributions of dust exposed to different radiation fields and temperatures.

\item Rotational disruption is the dominant dust-destruction mechanism on the considered timescales. In the early, embedded snapshot S$_1$, the dense natal cloud absorbs and reprocesses much of the UV radiation, reducing the efficiency of RATs and making the resulting SEDs nearly independent of the adopted maximum tensile strength $\mathcal{S}_{\mathrm{max}}$. In the evolved snapshot S$_6$, UV radiation propagates farther through the low-density ionized cavity, allowing RATs to spin up and disrupt LGs more efficiently. The resulting redistribution of grain mass towards smaller sizes produces a modest change in thermal dust emission but a more pronounced dependence of the AME amplitude on $\mathcal{S}_{\mathrm{max}}$. Rotational disruption of rapidly rotating NGs additionally suppresses the high-frequency tail of the AME spectrum. The comparatively small changes in thermal emission and stronger changes in AME suggest that multi-frequency microwave observations may help constrain the tensile strength of the grains.

\item Other destruction processes, including thermal sputtering, sublimation, and Coulomb explosion, affect only a small fraction of the simulation cells under the conditions considered here and do not significantly modify the global SEDs.

\item Hot dust emission exhibits a persistent positive alignment with $N_{\mathrm{H}}$ across the considered spatial scales. AME shows a moderate alignment with $N_{\mathrm{H}}$ that decreases towards larger scales. Free-free emission is largely uncorrelated or anti-aligned with $N_{\mathrm{H}}$, reflecting its dependence on the ionized gas distribution and emission measure rather than on the total gas column density.

\item Free-free emission is strongly aligned with $N_{\mathrm{el}}$, with the correlation approaching unity on large scales. AME exhibits a weaker but increasing alignment with $N_{\mathrm{el}}$ towards larger scales, whereas hot dust emission shows weak alignment or anti-alignment. These differences emphasize that thermal dust emission depends primarily on the radiation field and dust-bearing gas distribution, whereas free-free emission directly traces ionized gas.

\end{itemize}

 \appendix

\begin{acknowledgements}
We thank the anonymous referee for constructive feedback
on the manuscript. This project was funded by
the European Research Council via the ERC Synergy Grant “ECOGAL” (project
ID 855130). S.R. and R.S.K. acknowledge financial support from the Heidelberg cluster of excellence (EXC 2181 - 390900948) “{STRUCTURES}”.  R.S.K. also is grateful for funding from the German Ministry for Economy and Energy (BMWE) in project ``MAINN'' (funding ID 50OO2206), and from DFG and ANR for project ``STARCLUSTERS'' (funding ID KL 1358/22-1). The project made use of computing resources provided by {\em The L\"{a}nd} through bwHPC and by DFG through grants INST 35/1134-1 FUGG and 35/1597-1 FUGG. Data are in part stored at SDS@hd supported through grants INST I35/1314-1 FUGG and INST 35/1503-1 FUGG.
\end{acknowledgements}

\bibliographystyle{aa}
\bibliography{./bibtex}

\newpage

\section{Electric Dipole Moment of {\sc Astrodust}}
\label{app:dipole_ad}
The {\sc Astrodust} model developed by \cite{Draine2021ApJ} does not itself provide a consistent dipole moment. This limitation arises from the model's composition, which includes amorphous silicates, carbonaceous material, and other mineral constituents, with a fraction $f_{\rm Fe}$ of pure iron. Following \cite{Hoang2016}, we assume that the intrinsic dipole moments of the individual components within a dust grain add in a random-walk fashion. However, \cite{Hoang2016} provide dipole moments for only two of the {\sc Astrodust} components, forsterite and ferrosilite. To estimate an effective dipole moment for {\sc Astrodust}, we supplement these values with those available in the literature or estimate them based on structural similarities between the materials. A complete list of the adopted {\sc Astrodust} parameters is provided in Table~\ref{tab:astrodust}.

Following the random-walk prescription for randomly oriented intrinsic dipole moments, we calculate the effective dipole parameter of {\sc Astrodust} by summing the mean-square dipole moments of its constituents. For a grain material containing $N_{\mathrm{i}}$ atoms of each constituent, the root-mean-square dipole moment is
\begin{equation}
    \mu_{\mathrm{int}}
    = \left(\sum_{\mathrm{i}} N_{\mathrm{i}}\beta_{\mathrm{i}}^2\right)^{1/2}
    = \beta_{\mathrm{AD}}\sqrt{N_{\mathrm{at}}},
\end{equation}
where $\beta_{\mathrm{i}}$ is the characteristic dipole parameter of constituent $\mathrm{i}$ and $N_{\mathrm{at}}=\sum_{\mathrm{i}} N_{\mathrm{i}}$ is the total number of atoms. Expressing the number of atoms in terms of the constituent volume fraction $f_{\mathrm{i}}$, density $\rho_{\mathrm{i}}$, and mean mass per atom $m_{0,i}$ yields
\begin{equation}
    \beta_{\mathrm{AD}}
    = \left[
    \frac{
        \displaystyle\sum_{\mathrm{i}}
        f_{\mathrm{i}} \frac{\rho_{\mathrm{i}}}{m_{0,i}} \beta_{\mathrm{i}}^2
    }{
        \displaystyle\sum_{\mathrm{i}}
        f_{\mathrm{i}} \frac{\rho_{\mathrm{i}}}{m_{0,i}}
    }
    \right]^{1/2}.
\end{equation}
The effective grain material density $\rho_{\mathrm{gr}}$ and mean atomic mass $m_0$ are evaluated as
\begin{equation}
    \rho_{\mathrm{gr}} = \sum_{\mathrm{i}} f_{\mathrm{i}}\rho_{\mathrm{i}},
\end{equation}
and
\begin{equation}
    m_0 = \frac{\displaystyle\sum_{\mathrm{i}} f_{\mathrm{i}}\rho_{\mathrm{i}}}
         {\displaystyle\sum_{\mathrm{i}} f_{\mathrm{i}}\rho_{\mathrm{i}}/m_{0,i}}\, ,
\end{equation}
respectively. For the adopted {\sc Astrodust} composition with $f_{\rm Fe}=0$, corresponding to the absence of metallic iron inclusions, we obtain
$\beta_{\mathrm{AD}} \simeq 4.18\times10^{-30}\ \mathrm{Cm}$,
$\rho_{\mathrm{gr}} \simeq 3481\ \mathrm{kg\,m^{-3}}$,
and $m_0 \simeq 21.6\,m_{\mathrm H}$.

\begin{table*}[h!]
\centering
\caption{{\sc Astrodust} material parameters for the composition with with a content of $f_{\rm Fe}=0.0$. Values marked by est. are model estimates rather than values tabulated in literature.
References: [1] \cite{DHZ_Olivine_Pyroxene}; [2] Mineral Database (MINDAT) \url{https://www.mindat.org/}, accessed 2026 [3] \cite{Hunt2013}; [4] \cite{CRCHandbook}; 
[5] \cite{Cuda2011Troilite} [13] \cite{Saunders2011}; [14] \cite{Pekdur2020}; [15] \cite{Nagarajan2014} : [16] \cite{Valencia2020} ; [17] \cite{Sharipov2013} [18]  \cite{DraineLazarian1998}. }
\label{tab:astrodust}

\adjustbox{max width=\textwidth}{
\begin{tabular}{|l|c|c|c|c|c|}
\hline
Material & Formula & $f_{\mathrm{i}}$ [\%] & $\rho$ [kg\,m$^{-3}$] & $\beta$ [C\,m] & $m_0$ [kg] \\
\hline
Forsterite  & Mg$_2$SiO$_4$ & 28.2 & 3270$_{\ [1,3]}$ & $4.63\times10^{-30}_{\ [16]}$ & $3.34\times10^{-26}$ \\
Fayalite    & Fe$_2$SiO$_4$ & 6.9  & 4390$_{\ [2,3]}$ & $3.35\times10^{-30}_{\ [13],\,{\rm est.}}$ & $4.83\times10^{-26}$ \\
Enstatite   & MgSiO$_3$     & 13.5 & 3200$_{\ [1,3]}$ & $8.14\times10^{-30}_{\ [13]}$ & $3.33\times10^{-26}$ \\
Ferrosilite & FeSiO$_3$     & 3.3 & 4000$_{\ [2,3]}$ & $6.34\times10^{-30}_{\ [13]}$ & $4.38\times10^{-26}$ \\
Iron        & Fe            & 0.0   & 7870$_{\ [4]}$   & $0.0$ & $9.27\times10^{-26}$ \\
Troilite    & FeS           & 5.1  & 4700$_{\ [5]}$   & $3.67\times10^{-31}_{\ [15]}$ & $7.30\times10^{-26}$ \\
Calcite     & CaCO$_3$      & 4.4 & 2710$_{\ [4]}$   & $1.0\times10^{-30}_{\ {\rm est.}}$ & $3.32\times10^{-26}$ \\
Alumina     & Al$_2$O$_3$   & 1.6 & 4020$_{\ [4]}$   & $1.67\times10^{-30}_{\ [17]}$ & $3.39\times10^{-26}$ \\
Quartz      & SiO$_2$       & 2.8 & 2650$_{\ [4]}$   & $0.0$ & $3.33\times10^{-26}$ \\
Hydrogenated amorphous carbon & a-C:H & 19.3 & 2200$_{\ [4]}$ & $6.95\times10^{-31}_{\ [18],\,{\rm est.}}$ & $1.99\times10^{-26}$ \\
Magnetite   & Fe$_3$O$_4$   & 14.9 & 5170$_{\ [4]}$   & $1.20\times10^{-30}_{\ [14]}$ & $5.49\times10^{-26}$ \\
\hline
\end{tabular}
}
\end{table*}

\section{Multiscale correlation analysis}
\label{app:multiscale}
We quantify the spatial coupling between two non-periodic, two-dimensional fields
$X(\vec{x})$ and $Y(\vec{x})$. In detail, $X(\vec{x})$ represents column densities in our analysis and $Y(\vec{x})$ a synthetic map of intensity of a distinct tracer where as $\vec{x}=(x,y)^T$ are the coordinates in 2D pixel space. Our goal is to determine (i) whether the fields are correlated, (ii) on which spatial scales this correlation is strongest, and (iii) where in the map the coupling occurs. To this end we employ a multiscale decomposition based on Gaussian bandpass filtering, followed by global and local correlation diagnostics (see e.g. \citealt{StarckMurtagh1994} and \citealt{Frick2001} for further details).

The raw maps are transformed to logarithmic space to reduce dynamic range via ${X \leftarrow \log_{\mathrm{10}} \left[\max(X,\ \epsilon)\right]}$ 
%and ${Y \leftarrow \log_{\mathrm{10}} \left[\max(Y,\ \epsilon)\right]}$, respectively, 
where $\epsilon$ is a small positive floor to prevent undefined values. The maps are then standardized to zero mean and unit variance, 
by ${ \tilde{X}(\vec{x})=\left[X(\mathbf{x})-\langle X\rangle\right]/\sigma_{\mathrm{X}} }$ 
%and $\tilde{Y}(\mathbf{x})=\left[Y(\mathbf{x})-\langle Y\rangle\right]/\sigma_Y$
where $\langle \ \rangle$ denotes an average over all $\mathbf{x}$ and $\sigma_{\mathrm{X}}$ is the corresponding standard deviation. The $\tilde{Y}(\vec{x})$ maps of emission are created in a similar manner. We note that, in the following we drop tildes and assume standardized inputs unless stated otherwise.

To isolate structures on characteristic spatial scales, we define a bandpass-filtered representation of a field $X$ via a Difference-of-Gaussians (DoG),
\begin{equation}
W_{\mathrm{X}}(\mathbf{x},\Delta l)= \left(G_{\sigma_{\mathrm{h}}}\ast X\right)(\mathbf{x})-\left(G_{\sigma_\ell}\ast X\right)(\mathbf{x}),
\label{eq:dog}
\end{equation}
where $G_{\sigma}$ is a 2D Gaussian smoothing kernel with width $\sigma$, the operator $\ast$ denotes convolution, and $\Delta l$ is the dimensionless scale with respect to some reference lengths $l_0$. For simplicity we use the length of one pixel in our high-resolution maps leading to $l_0 = 1.5 \times 10^{-2}\ \mathrm{pc}$. We adopt $\sigma_\ell=\Delta l$ with a fixed ratio ${ \sigma_{\mathrm{h}} = 2.1 \sigma_\ell }$.
Consequently,  Eq.~\ref{eq:dog} acts as a wavelet-like bandpass filter that emphasizes features with sizes roughly between $\sigma_\ell$ and $\sigma_{\mathrm{h}}$ \citep[see e.g.][]{GillHenriksen1990,StarckMurtagh1994}. 

To localize which regions contribute most strongly to the global alignment at each scale, we define a local alignment proxy \citealt{Frick2001},
\begin{equation}
L(\mathbf{x},\Delta l)=\frac{W_{\mathrm{X}}(\mathbf{x},\Delta l)\,W_{\mathrm{Y}}(\mathbf{x},\Delta l)}{\sigma_{W_{\mathrm{X}}(\Delta l)}\ \sigma_{W_{\mathrm{Y}}(\Delta l)}}\,,
\label{eq:local_proxy}
\end{equation}
where $\sigma_{W_{\mathrm{X}}(\Delta l)}$ denotes the standard deviation of the coefficient map $W_{\mathrm{X}}(\mathbf{x},\Delta l)$, and analogously for $W_{\mathrm{Y}}$. We note that the quantity $L(\mathbf{x},\Delta l)$ is not normalized to unity.

For each scale $\Delta l$, we compute the global alignment coefficient between the bandpass-filtered fields $W_{\mathrm{X}}(\mathbf{x},\Delta l)$ and $W_{\mathrm{Y}}(\mathbf{x},\Delta l)$,
\begin{equation}
r(\Delta l)=\frac{\left\langle W_{\mathrm{X}}(\mathbf{x},\Delta l)\,W_{\mathrm{Y}}(\mathbf{x},\Delta l)\right\rangle}
{\sigma_{W_{\mathrm{X}}(\Delta l)}\ \sigma_{W_{\mathrm{Y}}(\Delta l)}}\,.
\label{eq:r_of_a}
\end{equation}
The statistic satisfies $r(\Delta l)\in\ ]-1,1[$ and measures whether structures of characteristic size $\sim \Delta l$ in $X$ and $Y$ are aligned ($r(\Delta l)>0$), unrelated ($r(\Delta l)=0$), or anti-aligned  ($r(\Delta l)<0$) \citep[][]{Frick2001,ArshakianBeck2011}. We note that the spatial average of $L(\mathbf{x},\Delta l)$ equals $r(\Delta l)$, hence $L(\mathbf{x},\Delta l)$ can be interpreted as the per-pixel contribution to the global alignment at scale $\Delta l$ \citep[see e.g.][]{Frick2001}.

\section{List of physical quantities}
\label{app:quantities}
In table \ref{tab:params} we list all physical quantities used throughout this work and provides their definitions for reference.

\begin{table*}
\caption{Physical quantities used in this work.}
\label{tab:params}
\centering
\small
\begin{tabular}{ll|ll}
\hline\hline
Symbol & Description & Symbol & Description \\
\hline
$\alpha_{\mathrm{n}}$ & Neutral gas polarizability &
$\lambda_{\mathrm{peak}}$ & Peak AME wavelength \\

$\beta$ & Dipole parameter &
$\lambda_{\mathrm{pdt}}$ & Photodetachment threshold \\

$\chi$ & Charge ratio &
$L$ & Luminosity \\

$\Delta l$ & Characteristic length scale &
$L_\lambda$ & Spectral luminosity \\

$\epsilon$ & Charge offset parameter &
$m_{\mathrm{0}}$ & Mean atomic mass \\

$E_{\mathrm{C}}$ & Coulomb energy &
$m_{\mathrm{e}}$ & Electron mass \\

$E_{\mathrm{pdt}}$ & Photodetachment energy &
$m_{\mathrm{H}}$ & Hydrogen mass \\

$E_{\mathrm{pet}}$ & Photoelectric energy &
$m_{\mathrm{i}}$ & Ion mass \\

$F$ & Total damping coefficient &
$\mu$ & Mass-to-flux ratio \\

$F_{\mathrm{IR}}$ & IR damping coefficient &
$\mu_{\mathrm{ch}}$ & Charge dipole moment \\

$F_{\mathrm{ni}}$ & Neutral-ion damping coefficient &
$\mu_{\mathrm{dip}}$ & Total dipole moment \\

$F_{\mathrm{nn}}$ & Neutral-neutral damping coefficient &
$\mu_{\mathrm{int}}$ & Intrinsic dipole moment \\

$F_{\mathrm{p}}$ & Plasma damping coefficient &
$n(a)$ & Grain size distribution \\

$F_{\mathrm{pe}}$ & Photoelectric damping coefficient &
$n_{\mathrm{d}}$ & Total dust number density \\

$G$ & Total excitation coefficient &
$n_{\mathrm{el}}$ & Electron number density \\

$G_{\mathrm{IR}}$ & IR excitation coefficient &
$n_{\mathrm{g}}$ & Gas number density \\

$G_{\mathrm{ni}}$ & Neutral-ion excitation coefficient &
$n_{\mathrm{i}}$ & Ion number density \\

$G_{\mathrm{nn}}$ & Neutral-neutral excitation coefficient &
$n_{\mathrm{LG}}$ & Large grain number density \\

$G_{\mathrm{p}}$ & Plasma excitation coefficient &
$n_{\mathrm{NG}}$ & Nanograin number density \\

$G_{\mathrm{pe}}$ & Photoelectric excitation coefficient &
$n_{\mathrm{n}}$ & Neutral gas number density \\

$I_{\mathrm{gr}}$ & Grain moment of inertia &
$N_{\mathrm{at}}$ & Number of atoms in a grain \\

$I_\lambda$ & Spectral intensity &
$N_{\mathrm{el}}$ & Electron column density \\

$j_\lambda$ & Emission coefficient &
$N_{\mathrm{H}}$ & Gas column density \\

$J_\lambda$ & Mean radiation intensity &
$Q$ & Photon emission rate \\

$Q_{\mathrm{abs}}$ & Dust absorption efficiency &
$Q_{\mathrm{RAT}}$ & RAT efficiency \\

$Q_{\mathrm{sca}}$ & Dust scattering efficiency &
$R_{\star}$ & Stellar radius \\

$S$ & Electron sticking coefficient &
$S_\nu$ & Spectral flux density \\

$\mathcal{S}_{\mathrm{max}}$ & Maximum tensile strength &
$\sigma_{\mathrm{pdt}}$ & Photodetachment cross section \\

$T_{\mathrm{eff}}$ & Effective stellar temperature &
$T_{\mathrm{el}}$ & Electron temperature \\

$T_{\mathrm{g}}$ & Gas temperature &
$T_{\mathrm{rot}}$ & Grain rotational temperature \\

$T_{\mathrm{sub}}$ & Dust sublimation temperature &
$\tau$ & Reduced temperature \\

$\tau_{\mathrm{ed}}$ & Dipole damping timescale &
$\tau_{\mathrm{g}}$ & Gas drag timescale \\

$\tau_{\mathrm{IR}}$ & IR drag timescale &
$\tau_{\mathrm{sp}}$ & Sputtering timescale \\

$\tau_\lambda$ & Spectral optical depth &
$\tau_\nu$ & Spectral optical depth \\

$\tilde{J}$ & Radiation enhancement factor &
$u_\lambda$ & Spectral radiation energy density \\

$v_{\mathrm{th}}$ & Thermal velocity &
$\omega_{\mathrm{rot}}$ & Grain angular velocity \\

$\omega_{\mathrm{RD}}$ & Maximum rotational velocity &
$Y$ & Electron yield \\

$Y_{\mathrm{sp}}$ & Sputtering yield &
$Z_{\mathrm{gr}}$ & Grain charge \\

$Z_{\mathrm{i}}$ & Ion charge &
$Z_{\mathrm{lower}}$ & Lower grain charge limit \\

$Z_{\mathrm{max}}$ & Maximum grain charge &
$Z_{\mathrm{min}}$ & Minimum grain charge \\

$Z_{\mathrm{upper}}$ & Upper grain charge limit &
$\rho_{\mathrm{gr}}$ & Grain material density \\

\hline
\end{tabular}
\end{table*}

\end{document}